\documentclass[lettersize,journal]{IEEEtran}
\usepackage{amsmath,amsfonts}
\usepackage{algorithmic}
\usepackage{algorithm}
\usepackage{array}
\usepackage{textcomp}
\usepackage{stfloats}
\usepackage{url}
\usepackage{verbatim}
\usepackage{graphicx}
\usepackage[table]{xcolor}
\usepackage{xcolor} 
\usepackage{cite}
\usepackage{amssymb}   % For \mathbb
\usepackage{amsfonts}  % For \mathbb
\usepackage{booktabs}   % For professional-looking tables (\toprule, \midrule, \bottomrule)
\usepackage{bbding}
\usepackage{multirow}
\usepackage{subcaption} % 用于子图
\usepackage{caption}
\usepackage{CJKutf8}
\usepackage{tabularx}
\usepackage{array}
\usepackage{hyperref}
\usepackage{xr-hyper}

\begin{document}

% When Tone and Words Disagree: Towards Robust Speech Emotion Recognition under Acoustic-Semantic Conflict
\title{Robust Speech Emotion Recognition under Tone-Word Conflict: A Benchmark and Framework}

\author{Xiaojiang Peng$^\dagger$$^{*}$,~\IEEEmembership{Senior Member,~IEEE}, Dawei Huang$^\dagger$, Yongjie Lv, Ruijie Xiong, Chunxiang Jin, Bin Li, Xiaohui Wang, Zitong Yu,~\IEEEmembership{Senior Member,~IEEE}

\thanks{Dawei Huang and Xiaojiang Peng are with Shenzhen Technology University, Shenzhen, 518118, China (e-mail: huangdawei2023@email.szu.edu.cn; pengxiaojiang@sztu.edu.cn).} 
\thanks{Yongjie Lv, Ruijie Xiong, and Chunxiang Jin are with Ant Group, Hangzhou, China. Bin Li is with Skyworth Digital Technology Co., Ltd, Shenzhen, China. Xiaohui Wang is with Xiaopai Technology Co., Ltd, Shenzhen, China. Zitong Yu is with Great Bay University, Dongguan, China.} 

\thanks{$^\dagger$Equal Contribution.}
\thanks{$^*$Corresponding author.}
}

\markboth{Journal of \LaTeX\ Class Files,~Vol.~14, No.~8, August~2021}%
{Shell \MakeLowercase{\textit{et al.}}: A Sample Article Using IEEEtran.cls for IEEE Journals}


\maketitle

\begin{abstract}
Speech emotion recognition (SER) is a crucial component of human-computer interaction, attracting extensive attention from both industry and academia. However, existing SER systems typically assume alignment between vocal tone and lexical semantics, overlooking the real-world scenarios that involve tone-word conflict—-where the emotion conveyed by speech contradicts the literal meaning of the words. To bridge this gap, we introduce TWIN-SER (Tone-Word Incongruent SER), a benchmark for systematic evaluation under acoustic-semantic incongruence, and show that state-of-the-art models degrade severely under such incongruence.
To address this, we propose DAS (Disentangled Acoustic-Semantic fusion), a framework that mitigates tone-word conflict by explicitly disentangling acoustic and semantic pathways, selecting informative high-energy embeddings, and adaptively fusing them via a lightweight query-based attention mechanism. Specifically, DAS comprises three crucial modules: i) a heterogeneous feature extraction module that separately captures complementary acoustic and semantic representations from raw input; ii) a high-energy embedding selection module that identifies and retains the most discriminative embeddings; and iii) a Q-Former combination module that bridges the two pathways through cross-attention, enabling robust emotion prediction under incongruent conditions. Extensive experiments demonstrate that DAS consistently outperforms existing methods in tone-word conflict scenarios, as well as in standard in-domain and zero-shot settings. Our code and datasets are available at \url{https://github.com/24DavidHuang/FAS}.

% Speech Emotion Recognition (SER) systems often assume congruence between vocal emotion and lexical semantics. However, in real-world interactions, acoustic-semantic conflict is common yet overlooked, where the emotion conveyed by tone contradicts the literal meaning of spoken words. We show that state-of-the-art SER models, including ASR-based, self-supervised learning (SSL) approaches and Audio Language Models (ALMs), suffer performance degradation under such conflicts due to semantic bias or entangled acoustic–semantic representations. To address this, we propose the \textbf{Fusion Acoustic-Semantic (FAS)} framework, which explicitly disentangles acoustic and semantic pathways and bridges them through a lightweight, query-based attention module. To enable systematic evaluation, we introduce the \textbf{Conflict in Acoustic-Semantic Emotion} \textbf{(CASE)}, the first dataset dominated by clear and interpretable acoustic-semantic conflicts in varied scenarios. Extensive experiments demonstrate that FAS consistently outperforms existing methods in both in-domain and zero-shot settings.
% Notably, on the CASE benchmark, conventional SER models fail dramatically, while FAS sets a new SOTA with 59.38\% accuracy.
% Our code and datasets are available at \url{https://github.com/24DavidHuang/FAS}

\end{abstract}


\section{Introduction}
\IEEEPARstart{S}{peech} Emotion Recognition (SER), a fundamental task in affective computing, aims to automatically identify a speaker's emotional state from speech signals. Its applications span a wide range of domains, from enhancing user experience in virtual assistants to supporting emotional well-being in healthcare services~\cite{9543566, abdollahi2022artificial}. Traditional SER methods primarily rely on acoustic features such as filter banks or Mel-frequency cepstral coefficients (MFCCs)~\cite{wani2021comprehensive}, which predominantly capture paralinguistic attributes but lack rich semantic information, thereby yielding suboptimal performance in complex scenarios. Recent advances have shifted toward leveraging representations from speech–text pre-trained models and self-supervised learning (SSL) encoders~\cite{emotion2vec, Vesper, CLAP, hubert, whisper, Wavlm}, achieving state-of-the-art results on standard benchmarks such as IEMOCAP~\cite{IEMOCAP} and MELD~\cite{MELD}. 

However, this success is largely confined to scenarios of acoustic‑semantic congruence, where prosodic cues align with literal meaning—for instance, expressing "What a beautiful day!" in a joyful tone. In real‑world interactions, tone‑word conflict (i.e., acoustic‑semantic disagreement) is common yet overlooked: the emotion conveyed by the voice contradicts the literal meaning of the spoken words. Real‑world communication is replete with nuanced expressions like sarcasm or cold fury, where a speaker's true emotion—conveyed through acoustic cues—is decoupled from, or even antithetical to, the semantic content of their utterance. For example, upon learning that a colleague has been promoted, someone might say "Congratulations on your promotion!" in a flat or resentful tone, subtly revealing underlying envy rather than joy. In these prevalent yet challenging scenarios, the performance of current SER methods collapses.
\begin{figure}[t]
    \centering
    \includegraphics[width=\linewidth]{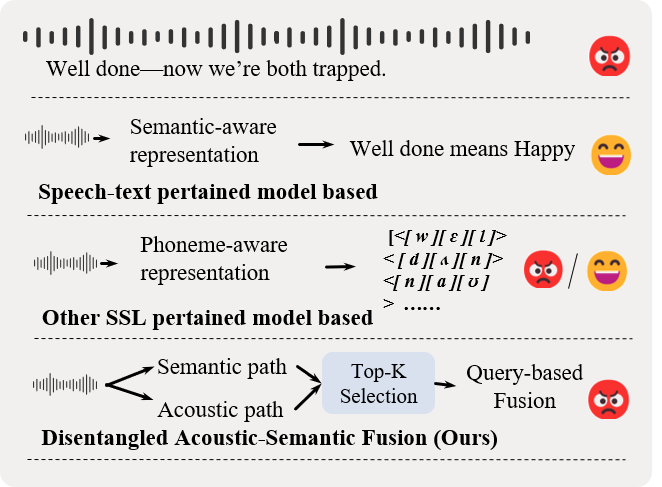}
    \caption{Comparison of SER methods on a conflicting utterance. Existing methods suffer from either semantic bias (poisoned by literal meaning) or entangled acoustic‑semantic representations.}
    \label{fig:moti}
    \vspace{-1.5em}
\end{figure}
% [现有sota方法对该问题的分析]-配图

This limitation is systemic across current state-of-the-art SER paradigms. As shown in Figure~\ref{fig:moti}, speech-text pre-trained model-based (e.g., Whisper~\cite{whisper} and CLAP~\cite{CLAP}) SER methods may exhibit semantic bias~\cite{10887715,10446997}, causing them to be ``poisoned'' by the literal meaning of the utterance. SSL pre-trained model-based (e.g., HuBERT~\cite{hubert}, wav2vec~\cite{wav2vec2.0}, and WavLM~\cite{Wavlm}) methods~\cite{DBLP:journals/corr/abs-2111-02735,DBLP:journals/corr/abs-2111-02735,jafarzadeh2024speaker} produce entangled representations where affective and semantic information are conflated, making ambiguity difficult to resolve. 
Recent audio-language large models (ALMs) such as Qwen2-Audio~\cite{qwen2_audio} and Qwen2.5-Omni~\cite{Qwen2_5_omni}, despite their impressive capabilities, depend on large language model (LLM)–aligned encoders that prioritize semantics over prosody, thereby depriving the downstream LLM of critical affective cues during emotional conflict.
All these structural brittlenesses limit the reliable deployment of SER models in tone-word conflict real‑world environments.

Beyond model-level deficiencies, the SER community also lacks a dedicated benchmark for systematic evaluation under tone-word conflict. Most benchmarks predominantly consist of congruent emotional expressions, where vocal prosody and lexical semantics align. Although in‑the‑wild datasets such as MELD~\cite{MELD} and IEMOCAP~\cite{IEMOCAP} contain sporadic instances of acoustic‑semantic conflict, these samples remain infrequent and lack structural control. Critically, no existing benchmark provides a high‑density, controlled setting for systematically evaluating SER robustness under emotional conflict—leaving a crucial dimension of real‑world performance unassessed.

To address these challenges, this work introduces a dual-pronged contribution. First, we propose \textbf{DAS (Disentangled Acoustic‑Semantic Fusion)} as shown in Figure \ref{fig:moti}, a framework that mitigates tone‑word conflict by explicitly disentangling acoustic and semantic pathways, selecting informative embeddings, and adaptively fusing them via a lightweight query‑based attention mechanism. DAS comprises three modules: a heterogeneous feature extraction module that separately captures low‑dimensional acoustic tokens (via an audio tokenizer) and high‑dimensional semantic representations (via a pre‑trained encoder); a high‑energy embedding selection module that identifies and retains the most discriminative embeddings while suppressing irrelevant information; and a Q‑Former combination module that bridges the two pathways through cross‑attention, dynamically weighting each modality to prioritize reliable cues under conflict. This design enables robust emotion prediction under tone‑word incongruence and strong generalization to zero‑shot SER benchmarks.

Second, to facilitate SER under acoustic‑semantic conflict and validate our proposed framework, we introduce \textbf{TWIN‑SER} (Tone‑Word Incongruent Speech Emotion Recognition Benchmark) for systematic evaluation. Unlike conventional datasets, TWIN‑SER is constructed with a high concentration of logical, interpretable, and scenario‑driven conflict samples. Specifically, we employ an LLM‑assisted approach for utterance generation and a state‑of‑the‑art text‑to‑speech model for audio generation. After manual verification, we obtain 378 high‑quality acoustic‑semantic conflict speech samples with multilingual and speaker‑diverse settings. TWIN‑SER serves not only as a challenging testbed for evaluating model robustness but also as a valuable corpus for studying the interplay between acoustics and semantics in human emotional expression.

The main contributions of this work can be summarized as follows:
\begin{itemize}
\item We are the first to systematically investigate the problem of acoustic–semantic conflict (tone–word conflict) in speech emotion recognition, revealing that existing state-of-the-art methods undergo significant performance degradation when confronted with such conflicting samples.
\item We introduce TWIN-SER, the first benchmark dataset specifically designed to evaluate model robustness against acoustic–semantic conflict, providing a controlled and high-density testbed for future research in this direction.
\item We propose DAS, a lightweight framework that explicitly disentangles acoustic and semantic features and adaptively fuses them via a query-based attention mechanism. Extensive experiments demonstrate that DAS consistently outperforms state-of-the-art baselines under conflict-centric settings, while achieving competitive performance on both in-domain and zero-shot benchmarks.
\end{itemize}

The remainder of this paper is organized as follows. Section \ref{sec:Related Work} reviews related work on speech emotion recognition, acoustic‑semantic conflict, and existing speech representation learning approaches. Section \ref{sec:method} details the proposed DAS (Disentangled Acoustic‑Semantic Fusion) framework, including its three core modules and training objective, followed by the TWIN‑SER benchmark. Section \ref{sec:exp} presents experimental results, including main comparisons, ablation studies, and zero-shot evaluations. Finally, Section \ref{sec:conclusion} concludes the paper with a summary of key findings, limitations, and directions for future work.

% To address these challenges, this paper introduces a dual-pronged contribution.
% First, we propose an innovative \textbf{Distangled Acoustic-Semantic Fusion (DAS)} framework, designed to explicitly disentangle acoustic and semantic information from speech. The FAS uniquely employs an audio tokenizer, inspired by recent advances in Text-to-Speech generation, to extract low-dimensional acoustic tokens, while concurrently utilizing a pre-trained module to capture high-dimensional semantic information.
% A lightweight, query-based module is then introduced to integrate disentangled features and make robust predictions.

% Second, to validate our proposed framework and to provide a much-needed resource for the community, we released the TWIN-SER (Tone-Word Incongruent
% SER), a benchmark for systematic evaluation. Unlike conventional datasets, CASE is constructed with a high concentration of logical, interpretable, and scenario-driven conflict samples. It serves not only as a challenging testbed for evaluating model robustness but also as a valuable corpus for researchers to study the interplay between acoustics and semantics in human emotion expression.
\section{Related Work}
\label{sec:Related Work}

\subsection{Speech Emotion Recognition}

Early speech emotion recognition (SER) research primarily relied on handcrafted acoustic features. Toolkits such as OpenSMILE~\cite{opensmile} were widely used to extract feature sets like eGeMAPS and ComParE, which comprise low-level descriptors (e.g., pitch, energy, and MFCCs) along with their statistical functionals. These features were subsequently fed into traditional machine learning classifiers, including support vector machines (SVMs) and Gaussian mixture models (GMMs)~\cite{Survey_on_SER}. With the advent of deep learning, the field shifted toward automatic feature learning, where Log-Mel spectrograms are commonly adopted as inputs to 2D convolutional neural networks (CNNs) to capture local time–frequency patterns~\cite{SER_CNN, SER_1D2D}.

More recently, large-scale self-supervised pre-trained models have significantly advanced SER performance. Encoders such as WavLM~\cite{Wavlm}, Whisper~\cite{whisper}, and HuBERT~\cite{hubert} are widely employed to extract high-level semantic–acoustic representations from their hidden layers, effectively alleviating the data scarcity problem in SER. Several studies~\cite{emotion2vec, Vesper, MFGCN} further focus on transferring, distilling, and adapting these powerful representations to downstream emotion recognition tasks.

Despite these advancements, core challenges remain unresolved. First, no universally optimal acoustic feature set has been established for reliably distinguishing emotional states, due to the high variability of speech signals across speakers, linguistic content, and speaking rates~\cite{Survey_on_SER}. Second, current pre-trained encoders introduce additional limitations: speech–text models such as Whisper~\cite{whisper} and CLAP~\cite{CLAP} exhibit strong semantic bias originating from their pre-training objectives, making them susceptible to literal meaning, while SSL-based encoders~\cite{hubert, Wavlm, wav2vec, wav2vec2.0} produce entangled representations where affective prosody remains inseparably conflated with phonetic content.

\subsection{SER in Acoustic-Semantic Conflict}
Research specifically targeting acoustic–semantic conflict in SER remains scarce. Existing related datasets, such as MUSTARD~\cite{MUSTARD}, provide direct supervision for conflict modeling (\textit{a binary classification problem}) by annotating inconsistencies between verbal and non-verbal cues. Similarly, iSarcasm and iSarcasmEval~\cite{iSarcasm/iSarcasmEval} focus on bilingual settings, revealing misalignments between cross-lingual prosody and literal meaning. However, these resources suffer from notable limitations. Their small scale and low ecological validity restrict generalizability; for instance, MUSTARD contains only 690 sitcom clips characterized by stylized performances that differ substantially from natural conversational speech. Furthermore, the use of coarse, subjective segment-level labels—often lacking contextual or causal grounding—hinders fine-grained conflict analysis and interpretable modeling.

To address these gaps, we propose a novel fusion framework that explicitly mitigates representation entanglement and semantic bias. In addition, we introduce the TWIN-SER benchmark, which is specifically designed for high-density acoustic–semantic conflict evaluation, providing a more controlled and challenging testbed for future research.

\subsection{Neural Audio Tokenization}

Advances in text-to-speech synthesis and audio editing have spurred the development of high-fidelity neural audio tokenizers. Discrete audio tokenizers, such as EnCodec~\cite{Encodec}, XCodec~\cite{Xcodec, Xcodec2}, and VibeVoice~\cite{VibeVoice}, are built upon the VQ-VAE framework~\cite{VQ-VAE} and excel at discretizing waveforms for high-quality signal reconstruction. More recently, VAE-based continuous tokenizers have been proposed to better unify semantic and acoustic information for joint understanding and generation tasks~\cite{Ming-UniAudio, DiTAR}, exemplified by MingTok-Audio~\cite{Ming-UniAudio}. A common characteristic of these generation-oriented approaches is their ability to distill speech into low-dimensional, acoustically clean representations. Unlike the high-dimensional, semantically entangled features produced by recognition-oriented encoders, these compact representations explicitly capture fine-grained prosodic and speaker-specific attributes, which are crucial for high-quality synthesis.

While neural audio tokenizers have primarily been applied to generative tasks, their potential as a source of disentangled acoustic representations for discriminative tasks such as SER remains largely unexplored. In this work, we pioneer the repurposing of audio tokens as a dedicated pathway for modeling acoustic affective features, aiming to resolve emotional ambiguity in scenarios where conventional SER methods are prone to failure.

\section{Methodology}
\label{sec:method}

\subsection{Overview}
State-of-the-art speech emotion recognition (SER) systems built upon self-supervised learning pre-trained speech models exhibit a fundamental vulnerability: they are predominantly biased toward either semantic content or paralinguistic cues, rendering them ineffective when these two modalities convey conflicting emotional signals—a common occurrence in real-world human communication known as tone–word conflict. To address this critical limitation and achieve robust performance under acoustic–semantic incongruence, we introduce DAS (Disentangled Acoustic–Semantic Fusion), a novel framework designed to explicitly separate, prioritize, and intelligently integrate information from distinct acoustic and semantic pathways. Moreover, we also present TWIN-SER (Tone–Word Incongruent Speech Emotion Recognition Benchmark), a dedicated evaluation benchmark engineered to systematically assess model robustness in the presence of such conflicts.

% State‑of‑the‑art SER methods based on pre‑trained models are primarily biased toward either semantic or paralinguistic representations, leaving them vulnerable to tone‑word conflict. To achieve robust speech emotion recognition under such conflict, we propose DAS (Disentangled Acoustic‑Semantic Fusion), a framework that explicitly disentangles acoustic and semantic pathways, selects high‑energy embeddings, and adaptively fuses them via a query‑based attention mechanism. Moreover, we introduce TWIN‑SER (Tone‑Word Incongruent Speech Emotion Recognition Benchmark), a conflict‑centric benchmark for systematic evaluation under acoustic‑semantic incongruence.

\begin{figure*}[t]
\centering
  \includegraphics[width=0.95\linewidth]{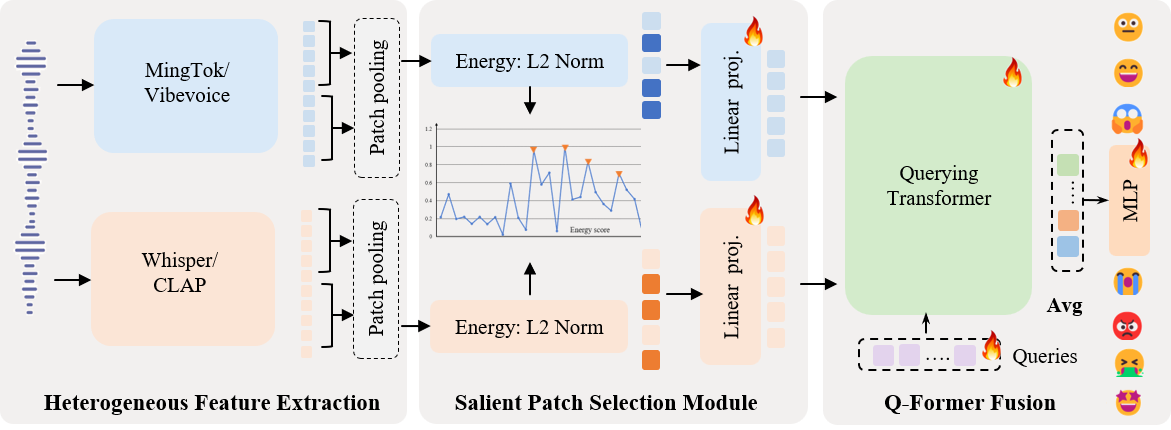}
  \caption{Overall framework of DAS (Disentangled Acoustic-Semantic Fusion). The model disentangles speech into two pathways: an acoustic representation (via MingTok-Audio) and a semantic representation (via Whisper). This disentanglement is crucial for handling tone-word conflict. It then selects high-energy patch embeddings and fuses them through a query-based Transformer for robust emotion recognition.}
  \label{fig:framework}
\end{figure*}

\subsection{Disentangled Acoustic-Semantic Fusion}

The overall framework of our DAS is illustrated in Figure~\ref{fig:framework}. DAS consists of three core components: a \textit{Heterogeneous Feature Extraction Module}, a \textit{Salient Patch Selection Module}, and a \textit{Q-Former Fusion Module}. 

\textbf{Heterogeneous Feature Extraction.}
We argue that both paralinguistic and linguistic information are essential for SER. Some speech pre-trained models are designed for ASR (e.g., Whisper~\cite{whisper} and CLAP~\cite{CLAP}), while others focus on speech generation and editing~\cite{Ming-UniAudio,Xcodec2,VibeVoice}. Our core insight lies in effectively fusing these two heterogeneous and temporally varying feature streams: the semantic features $F_{\text{sem}} \in \mathbb{R}^{T_{\text{sem}} \times D_{\text{sem}}}$ and the acoustic features $F_{\text{aco}} \in \mathbb{R}^{T_{\text{aco}} \times D_{\text{aco}}}$, where $T$ and $D$ denote sequence length and feature dimension, respectively. By default, we select MingTok~\cite{Ming-UniAudio} for paralinguistic features and Whisper for semantic features.

To efficiently handle long feature sequences, we apply a patchification step via average pooling over non‑overlapping windows of size $5$ following MingTok~\cite{Ming-UniAudio}, producing a shorter patch sequence $F' \in \mathbb{R}^{(T/5) \times D}$ for each pathway. 

\textbf{Salient Patch Selection.}
Recognizing that emotional cues are sparsely distributed, we introduce a non-uniform token selection strategy to identify and retain only the most informative ``highlight'' representations from each sequence. This is achieved as follows:
\begin{enumerate}
    \item Saliency scoring. We compute an energy score $s_t$ for each token $f_t$ in a sequence. Inspired by the observation that emotionally charged events often correlate with higher activation, we use the $\ell_2$ norm as a proxy:
    \begin{equation}
        s_t = \| f_t \|_2
        \label{eq:saliency_score}
    \end{equation}
    This fast, non-parametric method effectively captures moments of high energy in both the acoustic and semantic streams.
    
    \item Top-$k$ Selection. We then select the $k$ tokens with the highest saliency scores, where $k_{\text{aco}}$ and $k_{\text{sem}}$ are chosen to reflect the different information densities of each pathway. This yields two condensed sequences:
    \begin{equation}
        \begin{aligned}
            \mathbf{f}'_{\text{aco}} &\in \mathbb{R}^{k_{\text{aco}} \times D} ; \quad
            \mathbf{f}'_{\text{sem}} &\in \mathbb{R}^{k_{\text{sem}} \times D}.
        \end{aligned}
    \end{equation}
\end{enumerate}

This salient patch selection process drastically reduces sequence length while preserving the most emotionally relevant temporal information, offering an improvement over uniform compression techniques. 

Subsequently, these patch sequences are projected into a unified hidden dimension $d$: %=5
\begin{equation}
    \mathbf{f}_{\text{aco}} = \mathbf{f}'_{\text{aco}} W_{\text{aco}}; \quad  \mathbf{f}_{\text{sem}} = \mathbf{f}'_{\text{sem}} W_{\text{sem}}
\end{equation}
where $\mathbf{f}_{\text{aco}} \in \mathbb{R}^{K_{\text{aco}} \times d}$, $\mathbf{f}_{\text{sem}} \in \mathbb{R}^{K_{\text{sem}} \times d}$.

\textbf{Q-Former Fusion.} The selected sequence features are concatenated into a context sequence $\mathbf{f}_c' \in \mathbb{R}^{(k_{\text{aco}} + k_{\text{sem}}) \times d}$. We then employ a fusion module inspired by the Q-Former architecture~\cite{BLIP-2}. A set of $n$ learnable queries $Q \in \mathbb{R}^{n \times d}$ actively interrogate this context to produce a learned representation. The context is projected to generate key ($K$) and value ($V$) matrices. A cross-attention mechanism followed by residual connections and a feed‑forward network computes the final fused tokens:
\begin{align}
    \text{AttnOut} &= \text{Attention}(Q, W_k\mathbf{f}_c', W_v \mathbf{f}_c') \\
    Z &= \text{LayerNorm}(Q + \text{AttnOut}) \\
    X &= \text{LayerNorm}(Z + \text{MLP}(Z))
    \label{eq:attention_fusion}
\end{align}
Finally, a simple multi‑layer perceptron (MLP) serves as the prediction head on top of the mean fused vector to produce emotion probabilities $P(y|X)$ across the seven emotion categories.

\subsection{The TWIN-SER Benchmark}

\begin{figure*}[t]
  \includegraphics[width=1.0\linewidth]{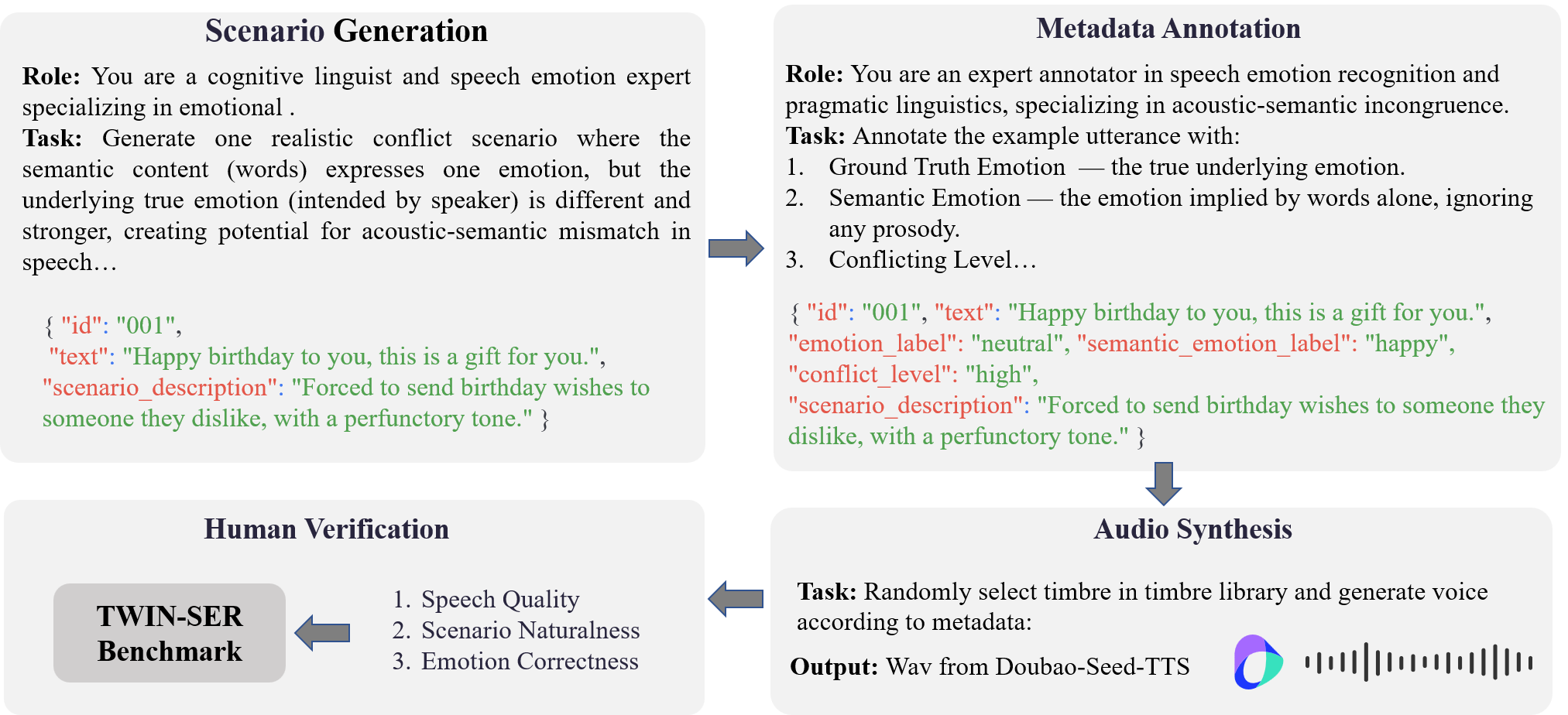}
  \caption{The four-stage construction pipeline of the TWIN-SER benchmark: scenario generation, metadata annotation, audio synthesis, and human verification.}
  \label{fig:twin}
\end{figure*}

To systematically assess model robustness under tone-word conflict, we present TWIN-SER (Tone–Word Incongruent Speech Emotion Recognition Benchmark), a dedicated evaluation benchmark engineered with a four-stage pipeline, as illustrated in Figure~\ref{fig:twin}. The construction process follows the principles of logical coherence and high conflict density, grounding every sample in plausible real-world scenarios with the assistance of an LLM.
In the first stage, Scenario Generation, we employ Gemini-2.5-pro~\cite{Gemini2.5} as a cognitive linguist and speech emotion expert to generate realistic conflict scenarios. Each scenario is designed such that the semantic content conveys one emotion while the underlying true emotion (specified via a scenario description) diverges, thereby creating an acoustic–semantic mismatch. For instance, the utterance “Happy birthday to you, this is a gift for you” is paired with the scenario: “Forced to send birthday wishes to someone they dislike, with a perfunctory tone.”
The second stage, Metadata Annotation, utilizes the same LLM to annotate each utterance with three key labels: the ground-truth emotion (the intended underlying emotion), the semantic emotion (inferred from the textual content alone), and the conflict level. In the above example, the ground-truth emotion is neutral, the semantic emotion is happy, and the conflict level is high.
In the third stage, Audio Synthesis, a speaker timbre is randomly sampled from a pool of 21 multi-emotion voices, and the corresponding audio is generated using the state-of-the-art TTS model Doubao-Seed-TTS 2.0~\cite{Seed-tts} based on the complete metadata.
Finally, Human Verification is conducted by a panel of 12 human experts, who manually evaluate each synthesized sample to ensure that the acoustic prosody clearly conveys the intended ground-truth emotion despite the semantic contradiction. Samples exhibiting weak, ambiguous, or semantically overshadowed prosody are discarded. This rigorous quality control process yields a final benchmark comprising 378 high-quality samples.

\textbf{TWIN-SER Dataset Statistics}.
The TWIN-SER benchmark comprises 378 high‑quality audio samples with 0.32 hours, synthesized using 21 distinct speaker timbres. The dataset exhibits a balanced gender distribution: 9 female speakers contribute 195 samples (51.59\%), while 12 male speakers account for 183 samples (48.41\%). In terms of language, the dataset is predominantly Chinese (230 samples, 60.85\%), with English constituting the remaining 148 samples (39.15\%).

Table~\ref{tab:distri} shows the distribution of ground-truth and semantic emotion labels across seven categories. The semantic labels, derived purely from the text, are predominantly neutral. In contrast, the ground-truth labels are more evenly distributed among the seven emotion classes, with no single category dominating the dataset. Representative samples from the TWIN-SER benchmark are provided in Table~\ref{tab:CASE_examples}.

\begin{table}[t]
\centering
\caption{Distribution of emotion labels in TWIN-SER: ground-truth labels vs. semantic-only labels.}
\label{tab:distri}
\resizebox{\linewidth}{!}{
\begin{tabular}{lccccccc}
\toprule
Label & sad & neutral & angry & happy & fear & surprised & disgust \\
\midrule
Ground-truth & 82 & 61 & 58 & 57 & 33 & 59 & 28 \\
Semantic     & 49 & 109 & 60 & 65 & 24 & 47 & 24 \\
\bottomrule
\end{tabular}
}
\end{table}

\begin{table*}[htbp]
\centering

\caption{Representative acoustic-semantic conflict samples from our TWIN-SER benchmark. For full audio demonstrations and additional metadata, please refer to the publicly released dataset files in the open-sourced repository.}
\begin{tabularx}{\textwidth}{
    l
    >{\raggedright\arraybackslash}p{5.2cm}
    c
    c
    >{\raggedright\arraybackslash}X
}
\toprule
ID & Utterance & GT Emo. & Semantic Emo. & Scenario Description \\
\midrule
% 001 & \begin{CJK}{UTF8}{gbsn}那辆卡车失控了，正朝我们冲过来！\end{CJK} & surprised & fear & An adrenaline-seeking extremist expresses morbid excitement at danger. \\
1 & \begin{CJK}{UTF8}{gbsn}你们又赢了，恭喜啊。\end{CJK} & angry & happy & A loser congratulates the winner with barely concealed resentment. \\
2 & \begin{CJK}{UTF8}{gbsn}他走了，再也不会回来了。\end{CJK} & happy & sad & Someone oppressed for years feels secret joy at their tormentor’s departure. \\
% 004 & \begin{CJK}{UTF8}{gbsn}你给我站住！你到底想怎么样！\end{CJK} & sad & angry & Exhausted from arguing; anger has turned into heartbreak and despair. \\
2 & \begin{CJK}{UTF8}{gbsn}任务完成，目标已清除。\end{CJK} & sad & neutral & A hitman reports completing a mission, but the target was an old friend. \\
4 & Well done—now we’re both trapped. & angry & happy & The speaker sarcastically blames their companion whose reckless actions led to a shared predicament, masking frustration with ironic praise. \\
5 & Mom, Dad... I love you. & fear & happy & A soldier records a final message to his parents before a suicide mission; his voice trembles with terror despite the loving words. \\
6 & No way—he was already dead! & fear & surprised & In a horror scenario, the protagonist witnesses a supposedly slain villain rise again, reacting with visceral fear beneath an initial gasp of shock. \\
7 & The emergency exit is blocked. & fear & neutral & During a fire, someone announces the only escape route is sealed—their tone calm in wording but laced with palpable panic and dread. \\
% 010 & What? You actually served this? & hate & surprised & A fastidious food lover reacts to a revolting “gourmet” dish with immediate disgust, their shock quickly overtaken by intense loathing. \\
8 & So this is your final decision, then? & disgust & neutral & After hearing an utterly unreasonable choice, the speaker delivers a cold, detached confirmation that conveys silent contempt and resignation. \\
9 &By the way, the building is on fire. We should probably leave. & neutral & fear &A character with a dry, British sense of humor and extreme stoicism delivering urgent, life-threatening news in a casual, conversational tone. \\
% 013 & Oh, a surprise party for me? You shouldn't have. & angry&neutral&An introvert who hates surprises is trying to be polite, but their voice is filled with irritation and anger. \\
% 014 & I hate you! I never want to see you again! &sad&angry& Saying hateful words during a breakup, but the underlying emotion is one of heartbreak and sadness.\\
10 & I love it. Another spreadsheet.&sad&happy&An employee sarcastically commenting on being assigned more tedious work, their voice full of gloom.\\
% 016 &You lost the game. It's over.&happy&neutral&A game show host playfully and cheerfully announcing bad news to a contestant. \\
% 017 & And the winner is... not you. & happy & neutral &A game show host playfully and cheerfully announcing bad news to a contestant.\\
% 018 &Don't worry about the dishes, I'll just do them. Again.&angry&neutral&A classic passive-aggressive roommate situation. The words are seemingly helpful, but the tone is dripping with anger and resentment. \\
% 019 &I heard you got the promotion. I am so, so thrilled for you.&sad&excited&Congratulating a coworker who got the promotion they wanted. They are trying to be supportive, but their voice is filled with their own disappointment. \\
% 020 &You're getting so mad over this little game, it's actually adorable.&happy&angry&A friend playfully teasing and taunting another friend who is getting frustrated while playing a video game.\\

\bottomrule
\end{tabularx}
\label{tab:CASE_examples}
\end{table*}

% \begin{table}[htbp]
% \centering
% \caption{Statistics of the TWIN-SER dataset.}
% \label{tab:twin_stats}
% \small
% \begin{tabular}{l r r}
% \toprule
% \textbf{Attribute} & \textbf{Category} & \textbf{Count (\%)} \\
% \midrule
% Total samples & -- & 378 (100.00\%) \\
% \multirow{2}{*}{Speakers} & Female & 9 (42.86\%) \\
%  & Male & 12 (57.14\%) \\
% \multirow{3}{*}{Language} & Chinese & 230 (60.85\%) \\
%  & English & 148 (39.15\%) \\
%  & Declared-vs-actual mismatch & 45 (11.90\%)$^*$ \\
% \midrule
% \multicolumn{3}{l}{\textbf{Ground-truth emotion}} \\
%  & Sad & 82 (21.69\%) \\
%  & Neutral & 61 (16.14\%) \\
%  & Angry & 58 (15.34\%) \\
%  & Happy & 57 (15.08\%) \\
%  & Fear & 33 (8.73\%) \\
%  & Surprised & 30 (7.94\%) \\
%  & Excited & 29 (7.67\%) \\
%  & Hate & 28 (7.41\%) \\
% \multicolumn{3}{l}{\textbf{Semantic emotion}} \\
%  & Neutral & 109 (28.84\%) \\
%  & Happy & 64 (16.93\%) \\
%  & Angry & 60 (15.87\%) \\
%  & Sad & 49 (12.96\%) \\
%  & Surprised & 31 (8.20\%) \\
%  & Hate & 24 (6.35\%) \\
%  & Fear & 23 (6.08\%) \\
%  & Excited & 16 (4.23\%) \\
%  & Tension & 1 (0.26\%) \\
%  & Lovey-dovey & 1 (0.26\%) \\
% \midrule
% \multicolumn{3}{l}{\textbf{Conflict level}} \\
%  & High & 231 (61.11\%) \\
%  & Medium & 113 (29.89\%) \\
%  & Subtle & 20 (5.29\%) \\
%  & None & 12 (3.17\%) \\
%  & Low & 2 (0.53\%) \\
% \bottomrule
% \end{tabular}
% \smallskip
% \footnotesize{$^*$ Samples where the declared speaker language mismatches the actual language used.}
% \end{table}

% \input{dataset}

\section{Experiments}
\label{sec:exp}

\subsection{Experiment Setup}

% ----- The Corrected Table -----
% 确保导言区有 \usepackage{booktabs}
\begin{table}[t]
\centering
\caption{Overview of datasets used for training and evaluation (excluding TWIN-SER). \#Emo denotes the number of emotion classes; Utts. is the number of utterances; \#Hours is the audio duration. The first six datasets are used for training, the middle three for general testing, and the last two for zero-shot testing.}
% \small % 使用小一号的字体来缩减表格整体大小
% \setlength{\tabcolsep}{10pt} % 减小列间距 (默认是 6pt)，可以根据需要微调
\resizebox{\linewidth}{!}{
\begin{tabular}{l c c c c c c}
\toprule
% \multicolumn{7}{c}{\textbf{Train \& In-Domain Test Sets}} \\
% \midrule
\textbf{Dataset} & \textbf{\#Emo} & \textbf{Utts.} &\textbf{\#Hours} & \textbf{Train} & \textbf{Test} & \textbf{Lang} \\
\midrule

IEMOCAP \cite{IEMOCAP}  & 5     & 10,039    &   7.0   & \checkmark     & $\times$ & English \\
CMU-MOSEI \cite{CMU-MOSEI} & 7       & 5,239    & 9.3 & \checkmark         & $\times$ & English  \\
MER2024 \cite{MER2024} & 6     & 5,030    &   5.9   & \checkmark         & $\times$ &Multilingual       \\
\rowcolor{gray!20} MELD \cite{MELD}  & 7       & 13,847    & 12.2 & 11.2         & 1.0 & English  \\
\rowcolor{gray!20} RAVDESS \cite{RAVDESS} & 8       & 2452     &  2.8  & 2.3    & 0.5  &  English\\
\rowcolor{gray!20} ESD \cite{ESD}  & 5   & 17,500  &  29.0  &  23.7 & 5.3 & Multilingual\\
% \midrule
% \multicolumn{7}{c}{\textbf{Zero-Shot Test Sets}} \\
% \midrule
% \textbf{CASE} (Ours) & 7 & 378 &  0.32  & -     & \checkmark     &  Multilingual \\
\rowcolor{gray!50} Emo-Emilia \cite{C2SER}  & 7& 1400 &  3.29 & $\times$  & \checkmark & Multilingual \\
% EMOVO \cite{EMOVO}  & 7   & 588  & 0.51 & -    & \checkmark  &  Italian \\
\rowcolor{gray!50} EmoDB \cite{EmoDB} & 7   & 535   & 0.41  & $\times$ & \checkmark     &  German \\
\bottomrule
\end{tabular}
}
\label{tab:datasets}
\end{table}
% -----------------------------

\textbf{1) Datasets \& Evaluation Metrics}:
Table~\ref{tab:datasets} provides an overview of the datasets used in our experiments. To build a generalized model, we aggregate multiple open-source datasets into a large-scale, heterogeneous training corpus totaling over 66 hours. This corpus comprises MER2024~\cite{MER2024}, a multilingual video-based emotion recognition corpus; IEMOCAP~\cite{IEMOCAP}, a dyadic conversational dataset of naturalistic emotional speech; CMU-MOSEI~\cite{CMU-MOSEI}, a large-scale sentiment analysis corpus with diverse topics and speakers; MELD~\cite{MELD}, a TV dialogue dataset from Friends; RAVDESS~\cite{RAVDESS}, a database of acted emotional speech and song by 24 professional actors; and ESD~\cite{ESD}, a multilingual emotional speech dataset with 350 parallel utterances from 10 English and 10 Chinese native speakers. All training samples from these datasets are used for model optimization.

For evaluation, we adopt weighted accuracy (WA) and weighted F1 (WF1) as the primary metrics. We report results on the test sets of MELD, RAVDESS, and ESD for general in-domain evaluation. Additionally, we assess model robustness on our TWIN-SER benchmark to examine performance under tone-word conflict. We also conduct an extra zero-shot evaluation on the test sets of Emo-Emilia~\cite{C2SER} and EmoDB~\cite{EmoDB} to further test generalization capability.

\textbf{2) Implementation Details}:
% ----- Hyperparameter Table -----
\begin{table}[t]
\centering
\caption{Hyperparameter settings for the experiments. The Concat~and~Gated column specifies the shared setting for the corresponding fusion baselines used in the ablation study.}
\begin{tabular}{l c c}
\toprule
\textbf{Hyperparameters} & \textbf{DAS} & \textbf{Concat\&Gated} \\
\midrule
Hidden Dimension ($d$) & 512 & 512 \\
Query Length ($N_q$) & 2 & -\\
Dropout Rate & 0.4 & 0.4\\
\midrule
Optimizer & \multicolumn{2}{c}{AdamW} \\
Learning Rate & \multicolumn{2}{c}{$2 \times 10^{-4}$} \\
LR Schedule & \multicolumn{2}{c}{Cosine} \\
Weight Decay & \multicolumn{2}{c}{$1 \times 10^{-4}$} \\
Global Batch Size  & \multicolumn{2}{c}{2048} \\
Loss & \multicolumn{2}{c}{Cross-Entropy} \\
Epochs & \multicolumn{2}{c}{100} \\
Warmup Ratio & \multicolumn{2}{c}{0.05} \\
Sample Rate & \multicolumn{2}{c}{16000} \\
\bottomrule
\end{tabular}
\label{tab:hyperparameters}
\end{table}
% -----------------------------
Our experiments were conducted on 8 NVIDIA A6000 GPUs. A fixed random seed of 42 was used for all experiments to ensure reproducibility. For the Semantic Pathway, we used the encoder from the pre-trained Whisper-large~\cite{whisper} model to extract 1280-dimensional features. For the Acoustic Pathway, we employed the MingTok-Audio~\cite{Ming-UniAudio} tokenizer to extract 64-dimensional features. Our proposed DAS was then trained from scratch on these precomputed features. This design accelerates experimentation by decoupling the heavy feature extraction process from the training of the lightweight fusion module.

As detailed in Table~\ref{tab:hyperparameters}, the fusion module is configured with a unified hidden dimension of $d=512$. The default selected acoustic and semantic salient patches are 8 and 16, respectively. The entire model is trained end-to-end using the AdamW optimizer \cite{AdamW} with an initial learning rate of $2 \times 10^{-4}$ and a weight decay of $1 \times 10^{-4}$. 
We use a global batch size of $2048$ and train for $100$ epochs, with Cross-Entropy Loss as the optimization objective.

\subsection{Evidence of Degradation under Tone-Word Conflict}
\begin{figure}
\centering
\includegraphics[width=\linewidth]{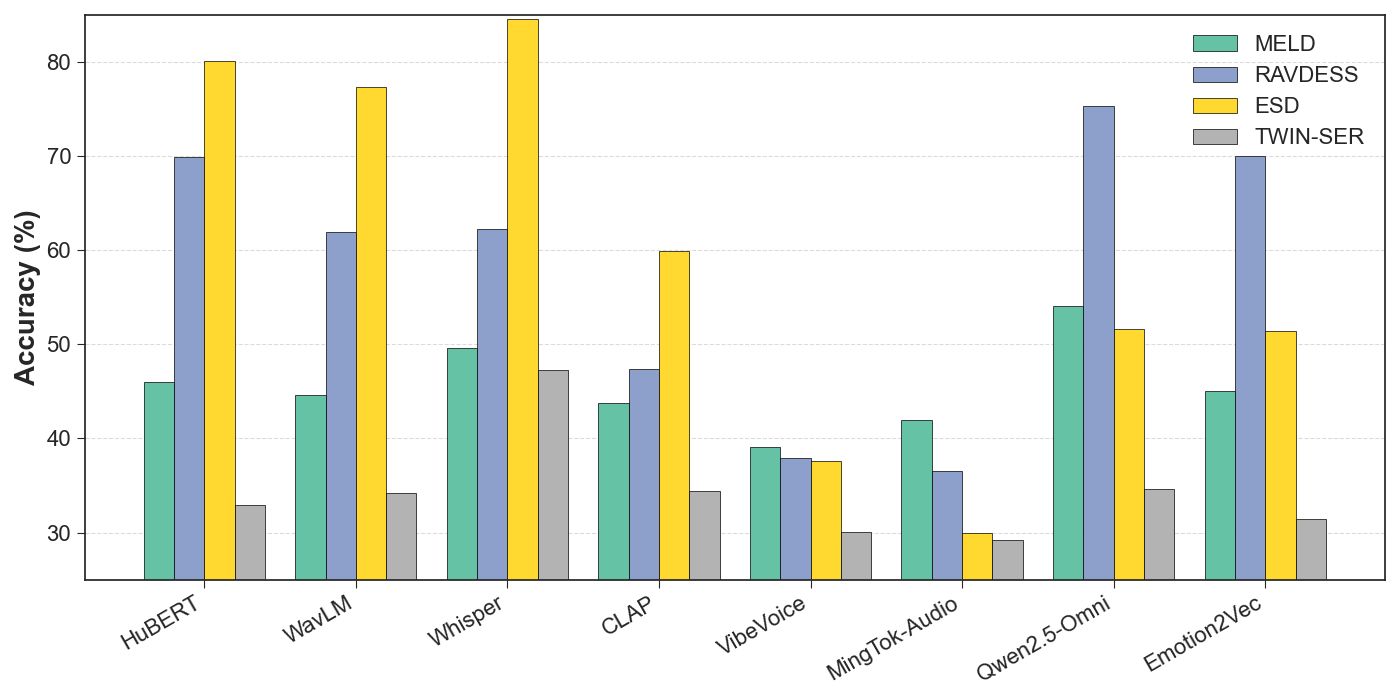}
\caption{Accuracy degradation of existing mainstream methods under tone-word conflict scenarios.}
\label{fig:evidence}
\end{figure}

We conduct comparative experiments covering eight representative speech emotion recognition approaches that span multiple paradigms: (1) self-supervised speech models (HuBERT~\cite{hubert}, WavLM~\cite{Wavlm}); (2) large-scale speech-text pre-trained encoders (Whisper~\cite{whisper}, CLAP~\cite{CLAP}); (3) neural audio tokenizers (VibeVoice~\cite{VibeVoice}, MingTok-Audio~\cite{Ming-UniAudio}); (4) a speech emotion dedicated model (Emotion2Vec~\cite{emotion2vec}); and (5)  general-purpose multimodal large-model audio encoders (Qwen2-Audio~\cite{qwen2_audio} and Qwen2.5-Omni 7B~\cite{Qwen2_5_omni}). 

To ensure a fair and rigorous comparison across these baselines, for all (1)--(4) type models, we freeze the pre-trained encoder, extract mean-pooled utterance embeddings, and train a lightweight two-layer classifier. Specifically, we extract the acoustic latent embeddings for MingTok-Audio. For Qwen2.5-Omni and  Qwen2-Audio, their outputs are projected onto a unified label space consistent with our target benchmarks for comparison.
Four standard speech emotion recognition datasets are adopted: MELD, RAVDESS, ESD, and our TWIN-SER, as shown in Figure~\ref{fig:evidence}. 

Several observations can be drawn as follows. First, all models except the neural audio tokenizers achieve strong performance on the three existing datasets. The underperformance of neural audio tokenizer models stems from their exclusive reliance on paralinguistic and prosodic features, with no capacity to leverage lexical semantics. Second, based on the results, we argue that acted datasets such as RAVDESS and ESD are more semantically biased, whereas MELD—a multi-modal emotion dataset—contains speech information that is only partially correlated with emotional expression. Most critically, all models exhibit dramatic performance degradation when evaluated on our TWIN-SER dataset. For instance, HuBERT achieves 80.1\% accuracy on ESD but drops sharply to only 32.9\% on TWIN-SER; similarly, Qwen2.5-Omni obtains 75.35\% on RAVDESS but falls to 34.66\% on TWIN-SER. These results clearly demonstrate that existing mainstream approaches severely fail in tone-word conflict scenarios.
\begin{table}[t]
\centering
\caption{Comparison with state-of-the-art methods. ``\(^{\dagger}\)''denotes results taken from the official papers. The best results are \textbf{bolded} and the second-best are \underline{underlined}.}
\label{tab:sota}
% \small
\setlength{\tabcolsep}{2.5pt}
\begin{tabular}{ l| cc  cc  cc cc }
\toprule
\multirow{2}{*}{\textbf{Methods}} 
& \multicolumn{2}{c}{\textbf{MELD}} 
& \multicolumn{2}{c}{\textbf{RAVDESS}} 
& \multicolumn{2}{c}{\textbf{ESD}} 
& \multicolumn{2}{c}{\textbf{TWIN-SER}} \\
& \textbf{WA} & \textbf{WF1} 
& \textbf{WA} & \textbf{WF1} 
& \textbf{WA} & \textbf{WF1} 
& \textbf{WA} & \textbf{WF1} \\
\midrule
HuBERT & 45.99&38.07&69.96&68.79&80.10&79.13&32.90&32.24\\
WavLM  & 44.64&34.09&61.90&60.69&77.43&76.63&34.20&33.92 \\
wav2vec2.0  & 37.44&28.36&23.59&16.44&21.81&19.57&25.59&22.83\\
% \midrule
\rowcolor{gray!20} Whisper  & 49.59&43.62&62.30&60.61&84.53&\underline{83.92}& \underline{47.26} & \underline{44.97} \\
\rowcolor{gray!20}CLAP & 43.74&34.96&47.38&44.83&59.97&59.40&34.46&31.93\\
EnCodec  & 34.38&29.41&22.78&18.44&28.73&25.18&24.54&18.81\\
Vibevoice & 39.06&30.43&37.90&31.48&37.61&36.40&30.03&25.90\\
MingTok-Audio & 41.94&29.76&36.49&26.12&29.93&28.80&29.24&25.61\\
\rowcolor{gray!20}Emotion2Vec\cite{emotion2vec} & 45.04 & 45.49 & 70.06 & 68.84 & 51.39 & 50.87 & 31.48 & 28.42 \\
\rowcolor{gray!20}Vesper~\cite{Vesper} \(^{\dagger}\) & 50.10  & 45.70 & - & - & - & - & - & - \\
\rowcolor{gray!20}$\text{C}^{2} \text{SER}$ \(^{\dagger}\)\cite{C2SER} & 51.39 & 27.45 & -&- & \textbf{93.86 }& 68.19 & -&- \\
\rowcolor{gray!20}TF-Mamba\(^{\dagger}\)\cite{10890265} &    & \textbf{48.5} & - & - & - & - & - & - \\
\rowcolor{gray!20}SeeNet\(^{\dagger}\)~\cite{li2025seenet} &    &  & \underline{80.5} & - & - & - & - & - \\
Qwen2-Audio & 35.29 & 29.91 &\textbf{85.74}&\textbf{86.59}& 36.99 & 23.35 &32.53&27.08 \\
Qwen2.5-Omni &\textbf{54.06} & 36.05 & 75.35 & 74.98 &51.60&35.70 & 34.66 & 30.21 \\
\rowcolor{gray!20}\textbf{DAS (Ours)} & \underline{51.89} &\underline{48.42}& 76.61 & \underline{76.19} & \underline{87.27} & \textbf{86.72} & \textbf{59.38} &\textbf{55.08}\\
\bottomrule
\end{tabular}
\end{table}

\subsection{Comparison with State-of-the-Art Methods}
We further compare our DAS against state-of-the-art SER methods with both accuracy and F1 score in Table~\ref{tab:sota}. The results consistently demonstrate that DAS achieves superior or competitive performance across all four datasets, with a particularly remarkable advantage on our proposed TWIN-SER, which features severe tone-word conflicts. On the three existing benchmarks (MELD, RAVDESS, ESD), existing methods exhibit highly variable performance. Overall, emotion-specific models tend to achieve the highest accuracy on one of the datasets, which can be attributed to their deliberately designed architectures tailored for emotion recognition.  Whisper achieves high accuracy on ESD (84.53\%) but drops substantially on MELD and RAVDESS. Qwen2.5-Omni, a large multimodal model, performs reasonably on RAVDESS (75.35\%) but only reaches 51.60\% on ESD and 34.66\% on TWIN-SER. In contrast, DAS delivers strong and balanced results: 87.27\% on ESD (second only to C$^2$SER's 93.86\%), 76.61\% on RAVDESS, and 51.89\% on MELD, while achieving state-of-the-art F1 scores on MELD (48.42\%), ESD (86.72\%), and TWIN-SER (55.08\%).

A notable observation can be made on TWIN-SER, where all baselines suffer severe degradation. Whisper, the best-performing baseline on this dataset, achieves only 47.26\% accuracy and 44.97\% F1. DAS substantially outperforms it by over 12\% and 10\% in F1, and nearly doubles the accuracy of HuBERT (32.90\%) and WavLM (34.20\%). Compared with emotion-specific methods, C$^2$SER~\cite{C2SER} achieves the highest accuracy on ESD (93.86\%) but its F1 is only 68.19\%, indicating potential overfitting or class imbalance, limiting its generalizability. Emotion2Vec, though self-supervised pre-trained on speech emotion datasets similar to ours, still underperforms on TWIN-SER. In contrast, DAS maintains a consistent accuracy–F1 balance across all datasets, confirming the effectiveness of our disentanglement and selective fusion strategies.
%that i) explicit disentanglement of tone and word emotional cues and ii) is critical for robust SER under conflicting scenarios.

% We evaluate FAS on both in-domain (MELD, RAVDESS, ESD) and zero-shot (CASE, Emo-Emilia, EMOVO, EmoDB) settings. As shown in Table~\ref{tab:comparison}, FAS achieves state-of-the-art results across the board: it obtains an average in-domain ACC of \textbf{71.92\%}, outperforming SSL models, semantic encoders, audio tokenizers, and even large audio-language models (ALMs) like Qwen2-Audio and Qwen2.5-Omni. Notably, while Qwen2.5-Omni excels on high-resource datasets (e.g., 87.85\% ACC on EmoDB), it underperforms on challenging zero-shot benchmarks such as CASE (34.66\%) and EMOVO (27.89\%). In contrast, FAS delivers robust performance—reaching \textbf{59.38\% SOTA ACC on CASE} and \textbf{54.66\% average ACC} across all zero-shot tasks—demonstrating its ability to generalize under distribution shift. 
% This performance gap reveals a fundamental trade-off in current ALMs: their architecture is optimized for alignment with the LLM backbone, which emphasizes textual semantics while drop the affective nuances carried by acoustic prosody.
% By explicitly modeling interactions between prosody and semantics, FAS bridges this gap, enabling reliable emotion recognition in both familiar and unseen scenarios.
% More experimental results including loss curves, confusion matrices, and visualizations of features map are provided in the \textit{Appendix Section~\ref{sec:additional_analyses}}.

\begin{figure}[t]
  \includegraphics[width=1.0\columnwidth]{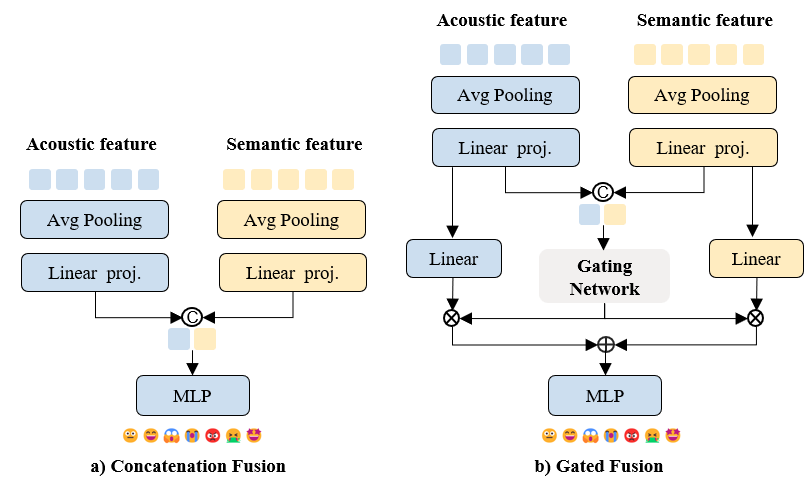}
  \caption{Illustration of different fusion strategies, including Concatenation Fusion and Gated Fusion.}
  \label{fig:fusion}
\end{figure}

\subsection{Ablation Study}
\label{sec:ablation_study}

To better understand the contributions of each component in our Disentangled Acoustic-Semantic (DAS) framework, we conduct ablation experiments addressing three key aspects: (1) the effectiveness of the proposed fusion and selection strategies; (2) the generalizability of our framework across different backbone encoders; and (3) the sensitivity to internal hyperparameters, specifically the number of selected salient patches and learnable queries.

\textbf{1) Effectiveness of Fusion and Selection Modules.} We first isolate the impact of the salient patch selection and the Q-Former fusion modules. For the selection strategy, we compare our energy-based top-$k$ selection against two alternatives: (i) random selection (Rand Sel.), which retains $k$ randomly chosen patches, and (ii) full attention (Full Att.), which applies standard multi-head self-attention over the complete patch sequence without any selection. These comparisons verify whether retaining temporally localized high-energy tokens is critical for capturing discriminative emotional bursts.  For the fusion architecture, as illustrated in Figure \ref{fig:fusion}, we replace our Q-Former with two representative alternatives: simple feature concatenation (Concat) and gated fusion (Gated). Specifically, the gated network employs a linear projection followed by a sigmoid activation to dynamically weight each modality. Both acoustic and semantic pathways are projected to a unified hidden dimension of $d=512$, consistent with our DAS configuration.

Table~\ref{tab:ablation_on_fusion} presents the results on MELD, RAVDESS, and our challenging TWIN-SER benchmark. For selection strategies, both random selection and full attention underperform our top-$k$ approach across all datasets, particularly on TWIN-SER, where DAS achieves 59.38\% accuracy versus 55.99\% for full attention and 55.47\% for random selection. This indicates that preserving salient, high-energy patches is essential for capturing emotionally charged moments, especially when prosodic and semantic cues conflict. For fusion strategies, Concat and Gated achieve comparable performance on MELD and RAVDESS, but lag behind DAS on TWIN-SER by approximately 5-6\% in accuracy. The Q-Former's learnable queries, which actively interrogate the context sequence, appear to better disentangle and re-integrate contradictory acoustic-semantic information, offering clear advantages in conflict-rich scenarios. Notably, although the full attention variant uses fewer parameters (0.82M), its inferior performance highlights the necessity of our selective and query-based design. Overall, these results confirm the effectiveness of both the salient patch selection and the Q-Former fusion modules, demonstrating their complementary roles in robust SER under tone-word conflicts.

\begin{table}[t]
\centering
\caption{SER performance with alternative fusion and selection strategies. All are built upon Whisper and MingTok-Audio encoders. Best results are \textbf{bolded} and the second-best are \underline{underlined}.}
 \resizebox{\linewidth}{!}{
\begin{tabular}{ l| c | cc  cc cc }
\toprule
\multirow{2}{*}{\textbf{Methods}} 
& \multirow{2}{*}{\textbf{Param}} 
& \multicolumn{2}{c}{\textbf{TWIN-SER}} 
& \multicolumn{2}{c}{\textbf{MELD}} 
& \multicolumn{2}{c}{\textbf{RAVDESS}} \\
& 
& \textbf{WA} & \textbf{WF1} 
& \textbf{WA} & \textbf{WF1} 
& \textbf{WA} & \textbf{WF1} \\
\midrule
Rand Sel.& 3.45M & 55.47&51.57 & 52.70&\textbf{48.60}&73.79&73.32\\
Full Att. & 0.82M & \underline{55.99}&\underline{52.48} &48.65&44.71&67.34&66.95\\
% Semantic Sel. & 3.45M & \underline{56.77}&\underline{53.86}&45.27&44.82&50.81&44.40 \\
% Acoustic Sel. & 3.45M & 28.65&25.80& 29.97&27.92& 36.49&26.12 \\
\rowcolor{gray!20}Concat & 1.22M & 53.65&50.77&\textbf{52.88}&\underline{48.50}&\underline{75.40}&\underline{75.22}\\
\rowcolor{gray!20}Gated  & 1.65M & 53.12&50.84& \underline{52.70}&47.92&73.79&73.59 \\
\rowcolor{gray!20}\textbf{DAS} (Ours)    & 3.45M &\textbf{59.38} &\textbf{55.08} &51.89&48.42 &\textbf{76.61}&\textbf{76.19}\\
\bottomrule
\end{tabular}
}
\label{tab:ablation_on_fusion}
\end{table}

\textbf{2) Generalizability across Backbone Encoders.} To assess whether DAS is compatible with other feature models (the default is Whisper for semantics and MingTok-Audio for acoustics), we systematically substitute each backbone with alternative pre-trained models. For the semantic pathway, we replace Whisper with CLAP~\cite{CLAP}, a cross-modal audio-text encoder. For the acoustic pathway, we substitute MingTok-Audio with VibeVoice~\cite{VibeVoice} and XCodec2~\cite{Xcodec2}, both of which are neural audio tokenizers originally designed for speech generation and editing.
\begin{table}[t]
\centering
\caption{Evaluation with diverse acoustic and semantic encoders for our DAS. The best results are \textbf{bolded} and the second-best are \underline{underlined}. }
\begin{tabular}{c c | c c c}
\toprule
\multirow{2}{*}{\textbf{Acoustic}}  & \multirow{2}{*}{\textbf{Semantic}}  & \textbf{TWIN-SER} & \textbf{MELD} & \textbf{RAVDESS} \\
& & WA / WF1& WA / WF1& WA / WF1\\
\midrule
- & CLAP& 34.46/31.93 &43.74/34.96&47.38/44.83 \\
MingTok & CLAP   &33.85/31.03 & 40.83/36.06 & 62.50/62.04\\
Vibevoice & CLAP& 36.72/34.19& 35.43/34.07&63.31/62.72\\
XCodec2 & CLAP & 32.55/30.63 & 43.26/38.14&59.07/58.81\\
% \midrule
\rowcolor{gray!20}- & Whisper & 47.26/44.97&49.59/43.62&62.30/60.61 \\
\rowcolor{gray!20}Vibevoice & Whisper  &58.07/53.35 & 51.53/48.06 & \textbf{80.04}/\textbf{79.71}\\
\rowcolor{gray!20}XCodec2 & Whisper & \underline{58.33/54.46}& \textbf{52.34}/\textbf{48.87}& \underline{79.03/78.76} \\
% \midrule
\rowcolor{gray!20}\textbf{MingTok} & \textbf{Whisper} & \textbf{59.38/55.08} &\underline{51.89/48.42} & 76.61/76.19 \\
\bottomrule
\end{tabular}
\label{tab:encoder_generalization}
\end{table}

Table~\ref{tab:encoder_generalization} presents the results on TWIN-SER, MELD, and RAVDESS. Several observations can be drawn. First, on acted emotion datasets such as RAVDESS, combining acoustic and semantic pathways substantially boosts performance; for example, CLAP+VibeVoice improves over CLAP alone by about 16\%, and Whisper+VibeVoice improves over Whisper alone by about 18\%. This underscores the importance of the two-pathway design. Second, when CLAP serves as the semantic encoder, its combinations with acoustic backbones yield significantly lower performance than those with Whisper, suggesting that CLAP may possess limited emotion perception ability acquired during pre-training and thus fails to capture fine-grained linguistic emotional cues, particularly in tone-word conflict scenarios. Third, with Whisper as the semantic encoder, pairing it with various acoustic encoders leads to consistent performance gains. Notably, XCodec2 paired with Whisper achieves 58.33\% WA and 54.46\% WF1 on TWIN-SER, closely approaching our default MingTok+Whisper configuration (59.38\%/55.08\%), while VibeVoice+Whisper attains the best results on RAVDESS (80.04\% WA). This consistency across different encoder pairs confirms that DAS effectively bridges heterogeneous acoustic and semantic representations, enabling robust cross-type fusion regardless of the specific backbone architectures. Finally, our default MingTok+Whisper combination achieves the best overall performance on TWIN-SER and competitive results on MELD and RAVDESS, validating our choice of encoders for the primary experiments. Overall, these results demonstrate that DAS is a generalizable framework that consistently improves SER performance across diverse encoder combinations, with its advantage being most pronounced on conflict-rich scenarios.

\begin{table}[t]
\centering
\caption{Performance comparison across varying numbers of selected patches for the acoustic ($k_{\text{aco}}$) and semantic ($k_{\text{sem}}$) pathways. The hidden dimension ($d=512$) and query count ($N_q=2$) remain constant throughout the experiments. The highest and second-highest scores for each dataset are denoted by \textbf{boldface} and \underline{underlining}, respectively.}
\begin{tabular}{cc  | cc cc cc }
\toprule
\multirow{2}{*}{\textbf{$k_{aco}$}} & \multirow{2}{*}{\textbf{$k_{sem}$}} 
& \multicolumn{2}{c}{\textbf{MELD}} 
& \multicolumn{2}{c}{\textbf{RAVDESS}} 
& \multicolumn{2}{c}{\textbf{TWIN-SER}} \\
& & \textbf{WA} & \textbf{WF1} 
& \textbf{WA} & \textbf{WF1} 
& \textbf{WA} & \textbf{WF1} 
\\
\midrule
8 & 8  & 51.62&47.99&77.82&77.51&\underline{58.59}&\underline{54.58}\\
8 & 16  & \underline{51.89}&\underline{48.42}&76.61&76.19& \textbf{59.38} &\textbf{55.08}\\
% \midrule
\rowcolor{gray!20}16 & 8  & 51.35&47.73&78.23&77.75&56.25&52.53\\
\rowcolor{gray!20}16 & 16  & 51.89&48.35&77.02&76.59&56.77&52.70\\
\rowcolor{gray!20}16 & 32  & \textbf{52.70}&\textbf{49.11}&\textbf{79.03}&\textbf{78.77}&56.25&52.37\\
% \midrule
32 & 16  & 50.90&47.44&\underline{78.83}&\underline{78.47}&55.21&51.47\\
\bottomrule
\end{tabular}
\label{tab:topK}
\end{table}
\textbf{3) The Number of Selected Salient Patches.} We systematically vary the number of selected patches for the acoustic pathway ($k_{\text{aco}}$) and the semantic pathway ($k_{\text{sem}}$) to analyze the trade-off between computational efficiency and recognition accuracy. The results are presented in Table~\ref{tab:topK}.  

From Table~\ref{tab:topK}, we observe that the optimal configuration varies across datasets. For MELD, the best performance is achieved with $k_{\text{aco}}=16$ and $k_{\text{sem}}=32$ (52.70\% WA, 49.11\% WF1), indicating a preference for richer semantic context in this multi-modal dialogue dataset. For RAVDESS, which relies heavily on exaggerated acted prosody and the semantics, the same configuration ($k_{\text{aco}}=16, k_{\text{sem}}=32$) yields the highest accuracy, suggesting that capturing sufficient acoustic details is particularly beneficial for acted emotional speech. For our challenging TWIN-SER benchmark, the optimal setting shifts to $k_{\text{aco}}=8$ and $k_{\text{sem}}=16$, achieving 59.38\% WA and 55.08\% WF1. Interestingly, using an equal number of patches (8 and 8) yields the second-best performance on TWIN-SER (58.59\% WA), while further increasing acoustic patches to 16 consistently degrades performance (e.g., 56.25\% WA for 16/8 and 56.77\% for 16/16). This suggests that in tone-word conflict scenarios, retaining too many acoustic patches may introduce misleading prosodic cues that interfere with lexical-semantic signals, whereas a compact acoustic selection combined with sufficient semantic patches effectively mitigates the conflict. Based on these observations, we adopt $k_{\text{aco}}=8$ and $k_{\text{sem}}=16$ as the default configuration for DAS, as it achieves the best robustness on the most challenging TWIN-SER benchmark while maintaining competitive performance on other datasets.

\begin{figure}[t]
  \includegraphics[width=1.0\columnwidth]{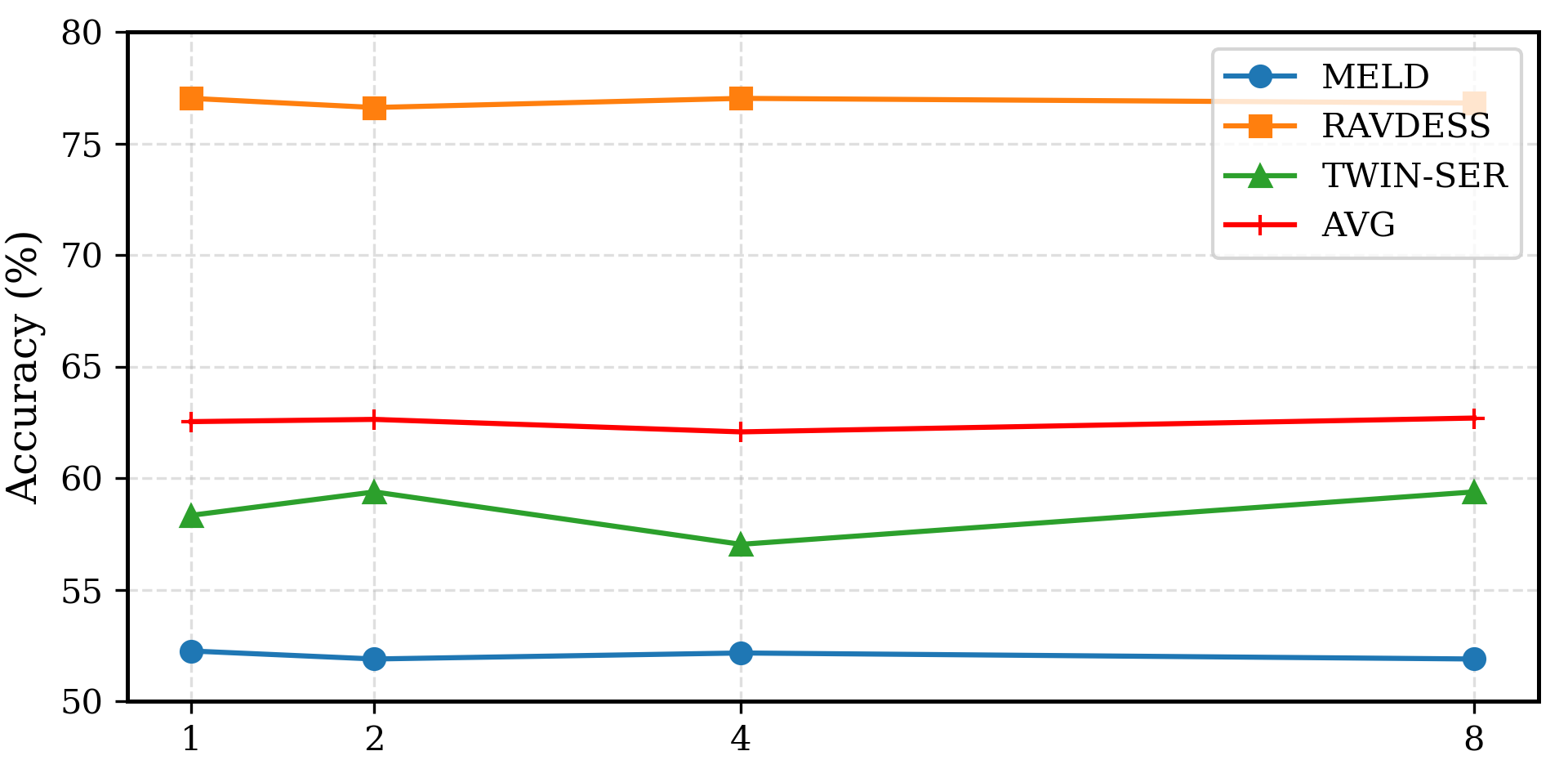}
  \caption{Ablation on the number of learnable queries ($N_q$). Average Accuracy is computed across the three datasets.}
  \label{fig:query_ablation}
\end{figure}

\textbf{4) The Number of Learnable Queries.} To assess the impact of the number of learnable queries ($N_q$) in the Q-Former fusion module, we experiment with $N_q \in \{1, 2, 4, 8\}$ while keeping all other hyperparameters fixed.
Figure~\ref{fig:query_ablation} reports the results on each dataset, alongside the average accuracy. On MELD and RAVDESS, the model exhibits robust performance with respect to $N_q$ across all configurations, suggesting that emotion cues can be integrated into a single dominant embedding on these non-conflict emotion datasets. performance on the conflict-rich TWIN-SER dataset shows noticeable variation depending on the number of queries, with the best result achieved at $N_q=8$ (60.5\% WA), followed by $N_q=2$ (59.5\%). The average accuracy across the three datasets remains stable with the best results at $N_q={2,8}$. Given that $N_q=2$ achieves highly competitive performance on TWIN-SER while maintaining identical performance on MELD and RAVDESS, and requires substantially fewer computations, we adopt $N_q=2$ as the default configuration for DAS to maximize both efficiency and robustness.

\subsection{Zero-Shot SER Evaluation}

In addition to the in-domain evaluation on MELD, RAVDESS, and ESD, we further assess the generalization capability of our DAS under zero-shot settings. Since all models are trained exclusively on the six-source training corpus without any exposure to our TWIN-SER benchmark, the results on TWIN-SER can naturally be viewed as a zero-shot evaluation. To provide a broader assessment of cross-dataset generalization, we additionally conduct experiments on two other unseen datasets: Emo-Emilia~\cite{C2SER} and EmoDB~\cite{EmoDB}, both of which contain acted emotional speech without deliberate tone-word conflicts. Table~\ref{tab:zero_shot_result} presents the comparison between DAS and mainstream approaches on TWIN-SER, Emo-Emilia, and EmoDB.

On our TWIN-SER benchmark, DAS achieves the best performance by a substantial margin, outperforming the strongest baseline Whisper by over 12\%. This confirms the effectiveness of our disentanglement and selective fusion strategies in handling tone-word conflict scenarios, even under zero-shot conditions. On Emo-Emilia and EmoDB, however, DAS achieves moderate results (51.14\%/42.92\% and 68.10\%/65.07\%, respectively), which are competitive with or exceed general-purpose models like Whisper, but fall behind task-specific (Emotion2Vec) or large multimodal approaches. Notably, Emotion2Vec achieves strong performance on EmoDB (71.21\% WA, 76.07\% WF1) and Emo-Emilia (52.79\% WA), benefiting from its self-supervised pre-training on large-scale emotion data. Meanwhile, Qwen2.5-Omni attains the highest scores on both Emo-Emilia (70.64\% WA) and EmoDB (87.85\% WA), reflecting the advantage of its extensive multimodal pre-training. 
Overall, these results reveal a clear trade-off: DAS is specifically designed to resolve tone-word conflicts, and thus excels on challenging benchmarks like TWIN-SER that exhibit such conflicts. On standard acted datasets without deliberate incongruence, dedicated emotion models or large foundation models with broader pre-training may offer advantages. Nevertheless, DAS demonstrates competitive zero-shot generalization across diverse datasets, and its superior performance on conflict-rich scenarios highlights its unique value for robust SER.% in real-world applications where acoustic-semantic mismatches are common.

\begin{figure*}[t]
\centering
\includegraphics[width=\textwidth]{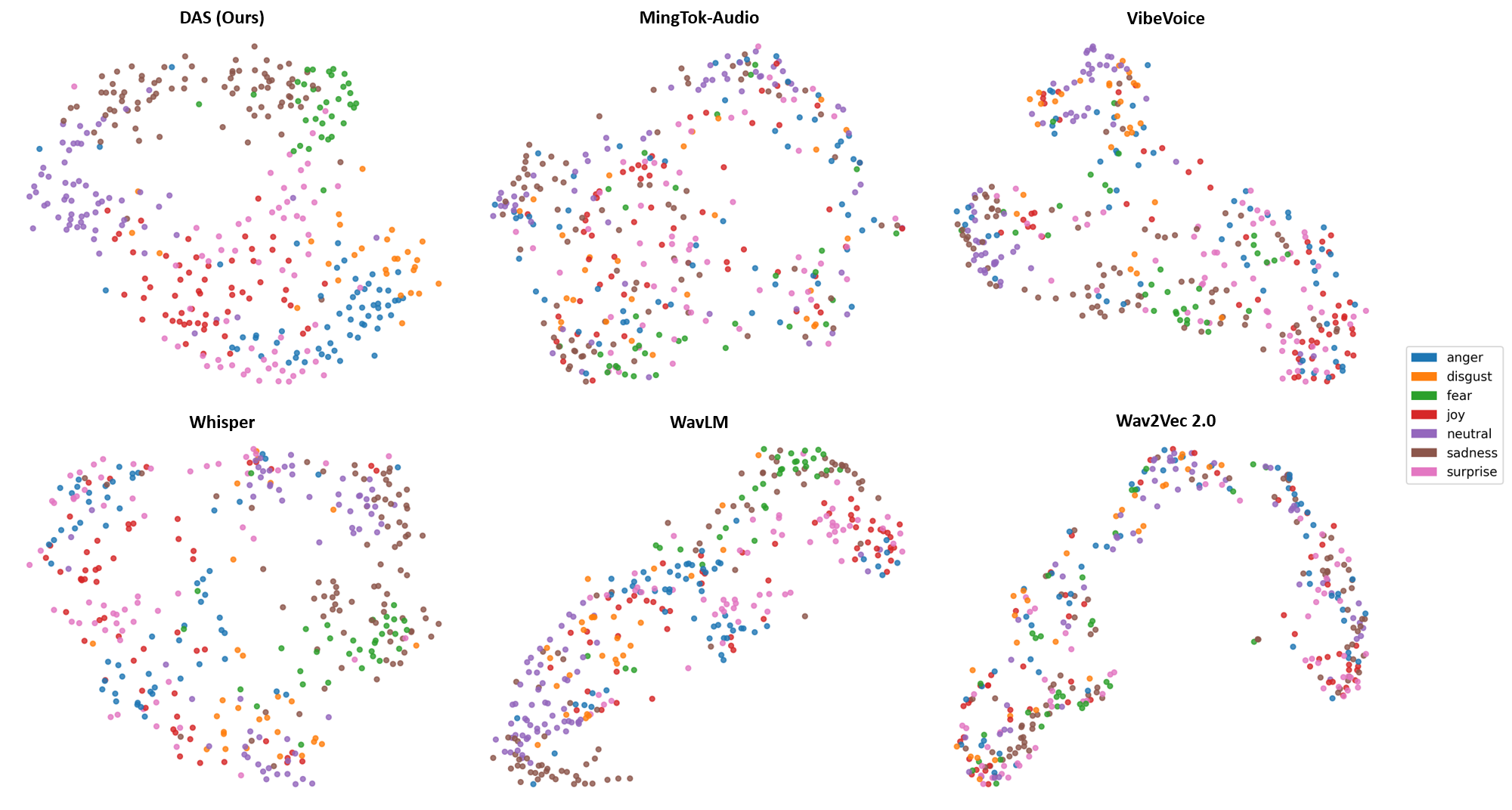}
\caption{UMAP visualization of representations on TWIN-SER benchmark. Colors denote ground-truth acoustic emotion labels. }
\label{fig:umap_CASE}
\end{figure*}
\subsection{Visualization}
\label{sec:vis}

\begin{table}[t]
\centering
\caption{Compared our DAS to mainstream approaches under a non-conflict zero-shot setting. The highest and second-highest scores for each dataset are denoted by \textbf{boldface} and \underline{underlining}, respectively.}
% \resizebox{\textwidth}{!}{
\begin{tabular}{l|cc|cc|cc}
\toprule
\multirow{2}{*}{\textbf{Methods}} &\multicolumn{2}{c|}{TWIN-SER} & \multicolumn{2}{c|}{Emo-Emilia} &  \multicolumn{2}{c}{EmoDB} \\
&WA & WF1 & WA & WF1 &  WA & WF1 \\
\midrule
HuBERT&32.90 & 32.24 & 34.36 & 29.95 &  52.99 & 53.57 \\
WavLM&34.20 & 33.92 & 35.64 & 29.33 & 64.18 & 56.74 \\
wav2vec2.0&25.59 & 22.83 & 19.64 & 16.02 &  25.37 & 22.09 \\
% \hline
\rowcolor{gray!20}Whisper&\underline{47.26} & \underline{44.97} & 50.50 & 42.29 & 68.84 & 63.46 \\
\rowcolor{gray!20}CLAP&34.46 & 31.93 & 24.93 & 18.83 & 43.28 & 41.45 \\
% \hline
EnCodec&24.54 & 18.81 & 15.64 & 11.47 &  29.10 & 23.87 \\
Vibevoice&30.03 & 25.90 & 19.36 & 15.35 & 28.73 & 20.92 \\
MingTok-Audio&29.24 & 25.61 & 21.07 & 16.34 &  37.31 & 34.02 \\
% \hline
\rowcolor{gray!20}Emotion2Vec&31.48 & 28.42 & 52.79 & 50.44 &  71.21 & \underline{76.07} \\
% \hline
% $\text{C}^{2} \text{SER}$&-- & -- & 68.29 & 61.28 & -- & -- \\
Qwen2-Audio&32.53 & 27.08 & \underline{69.64} & \textbf{68.81} & \underline{74.21} & 70.29 \\
Qwen2.5-Omni&34.66 & 30.21 & \textbf{70.64} & \underline{68.03} & \textbf{87.85} & \textbf{85.97} \\
\midrule
\rowcolor{gray!20}\textbf{DAS}(Ours)&\textbf{59.38} &\textbf{ 55.08} & 51.14 & 42.92 &  68.10 & 65.07 \\
\bottomrule
\end{tabular}
% }
\label{tab:zero_shot_result}
\end{table}

\begin{figure*}[htbp]
    \centering
    % 第一行（3张图）
    \begin{subfigure}[b]{0.30\textwidth}
        \includegraphics[width=\linewidth]{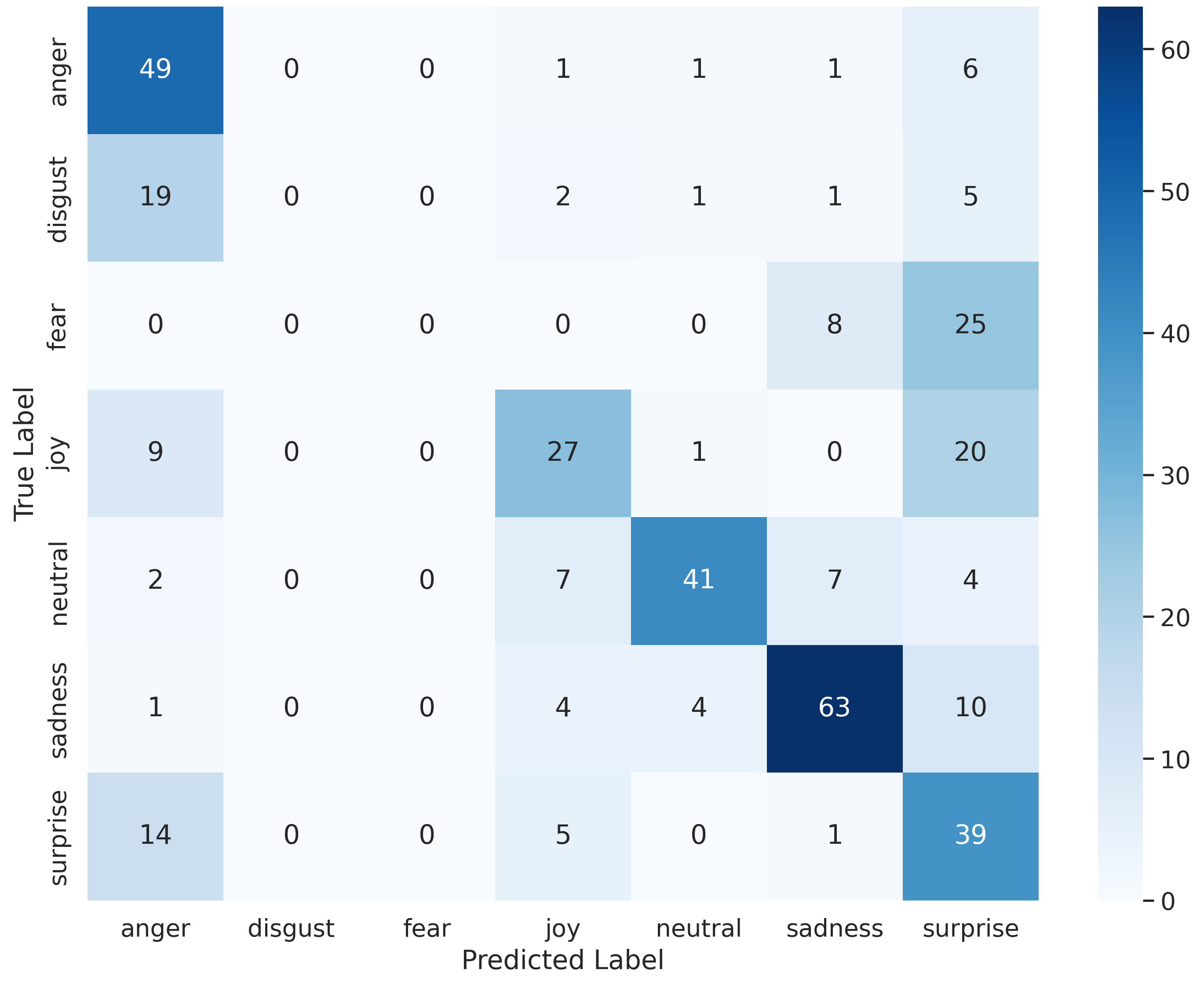}
        \caption{DAS (Ours)}
        \label{fig:CM_CASE_1a}
    \end{subfigure}
    \hfill
    \begin{subfigure}[b]{0.30\textwidth}
        \includegraphics[width=\linewidth]{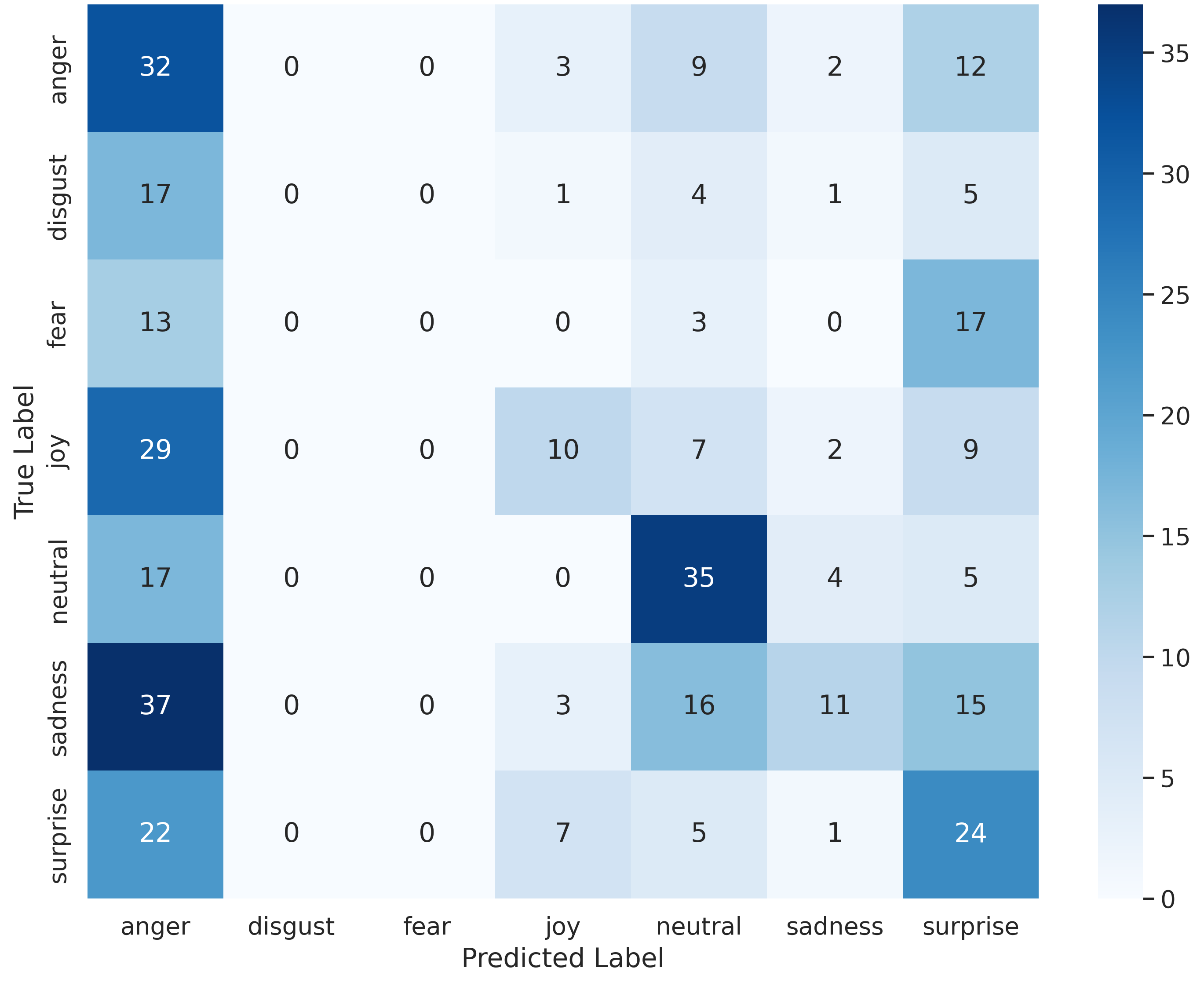}
        \caption{MingTok-Audio }
        \label{fig:CM_CASE_1b}
    \end{subfigure}
    \hfill
    \begin{subfigure}[b]{0.30\textwidth}
        \includegraphics[width=\linewidth]{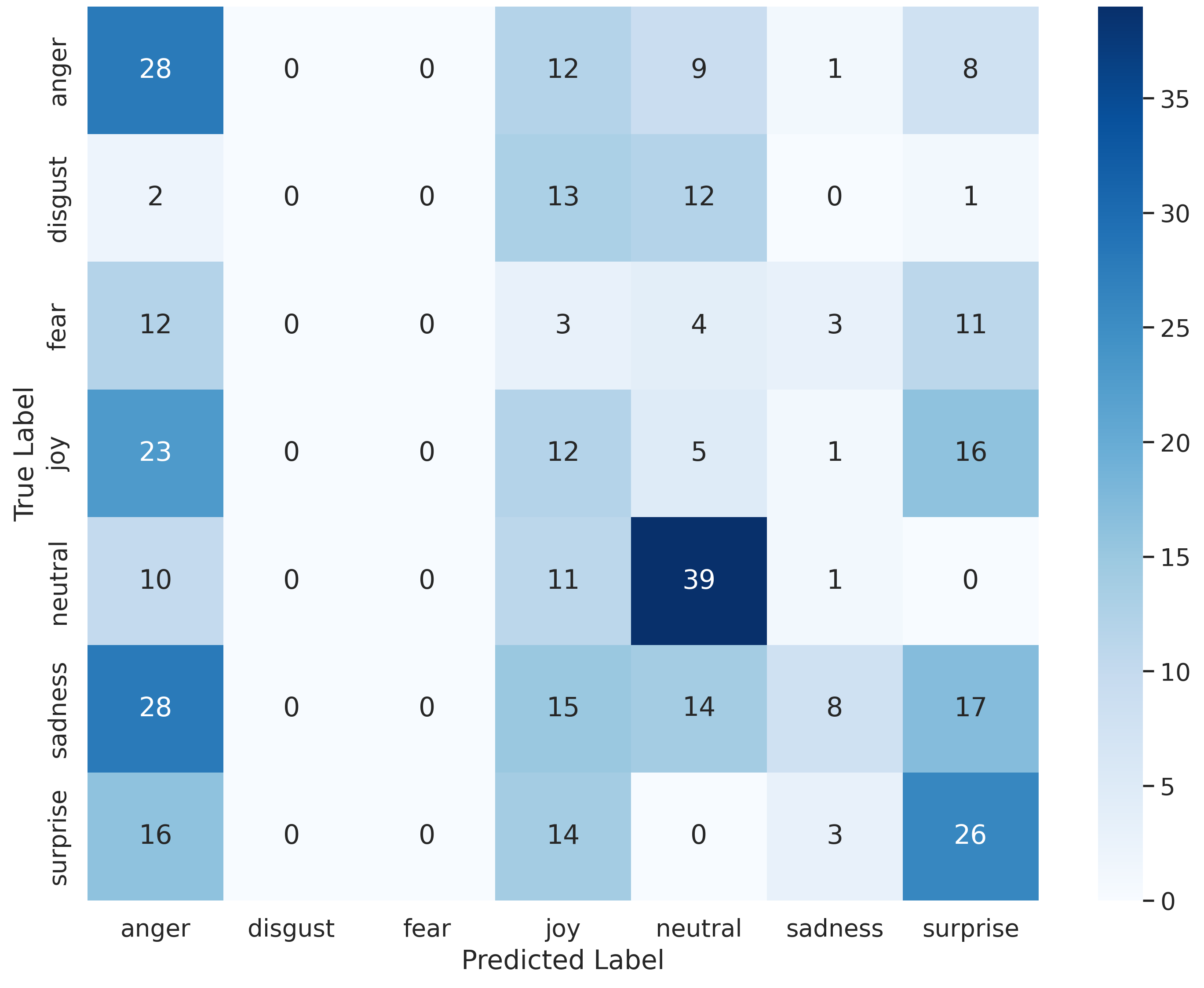}
        \caption{VibeVoice }
        \label{fig:CM_CASE_1c}
    \end{subfigure}

    % \\ % ← 关键！强制换行到第二行

    % 第二行（3张图）
    \begin{subfigure}[b]{0.30\textwidth}
        \includegraphics[width=\linewidth]{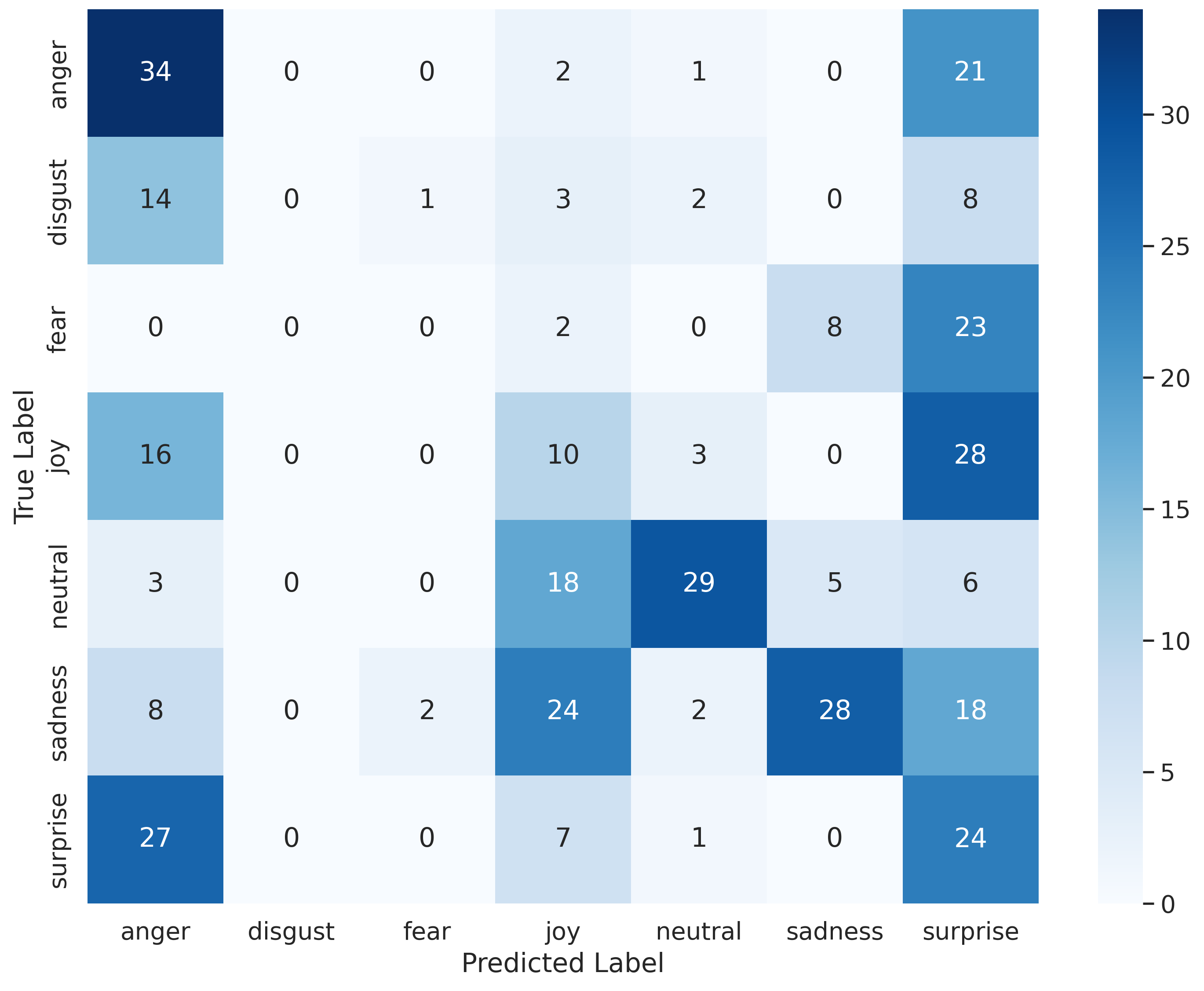}
        \caption{HuBERT} %
        \label{fig:CM_CASE_2a}
    \end{subfigure}
    \hfill
    \begin{subfigure}[b]{0.30\textwidth}
        \includegraphics[width=\linewidth]{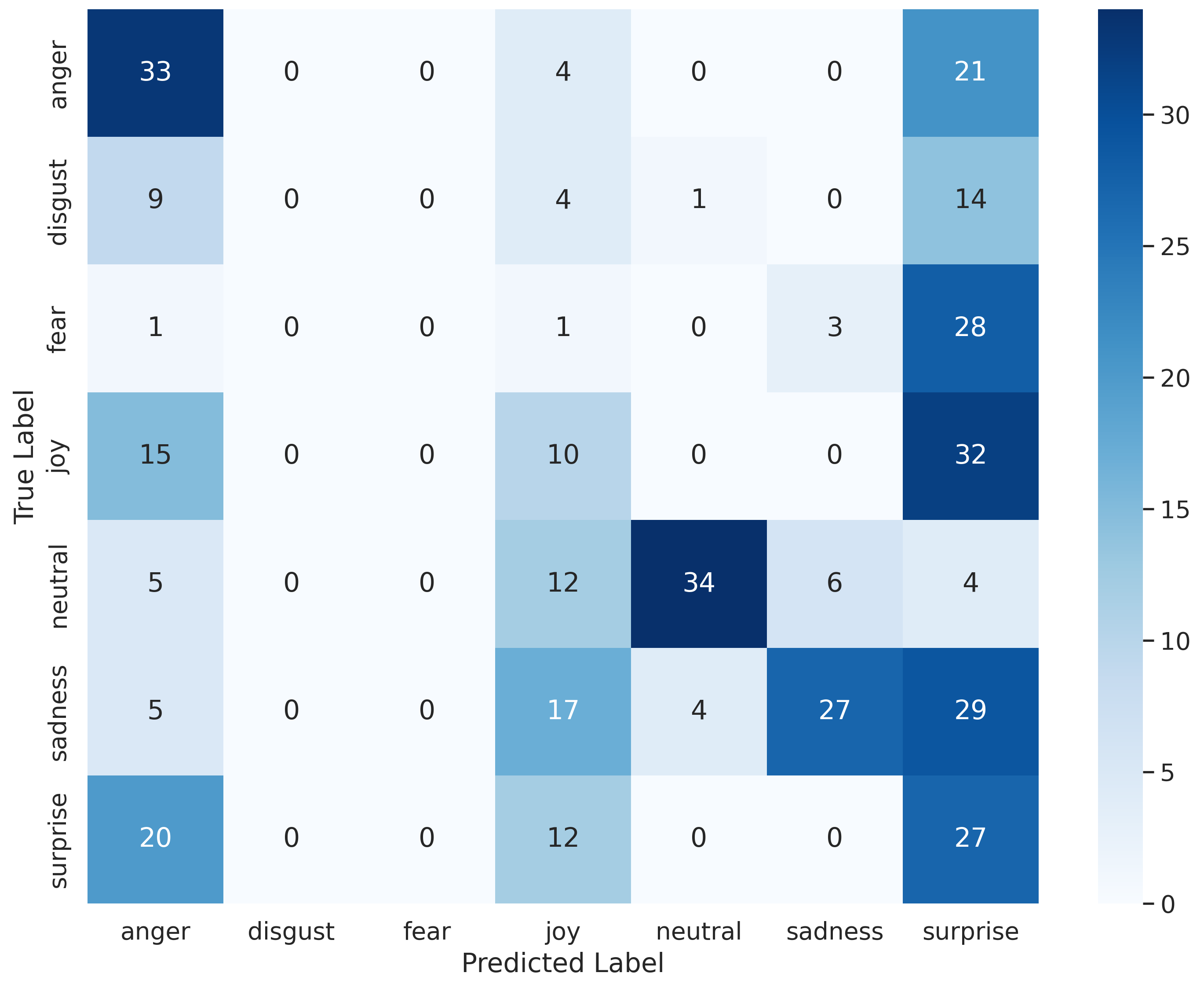}
        \caption{WavLM }
        \label{fig:CM_CASE_2b}
    \end{subfigure}
    \hfill
    \begin{subfigure}[b]{0.30\textwidth}
        \includegraphics[width=\linewidth]{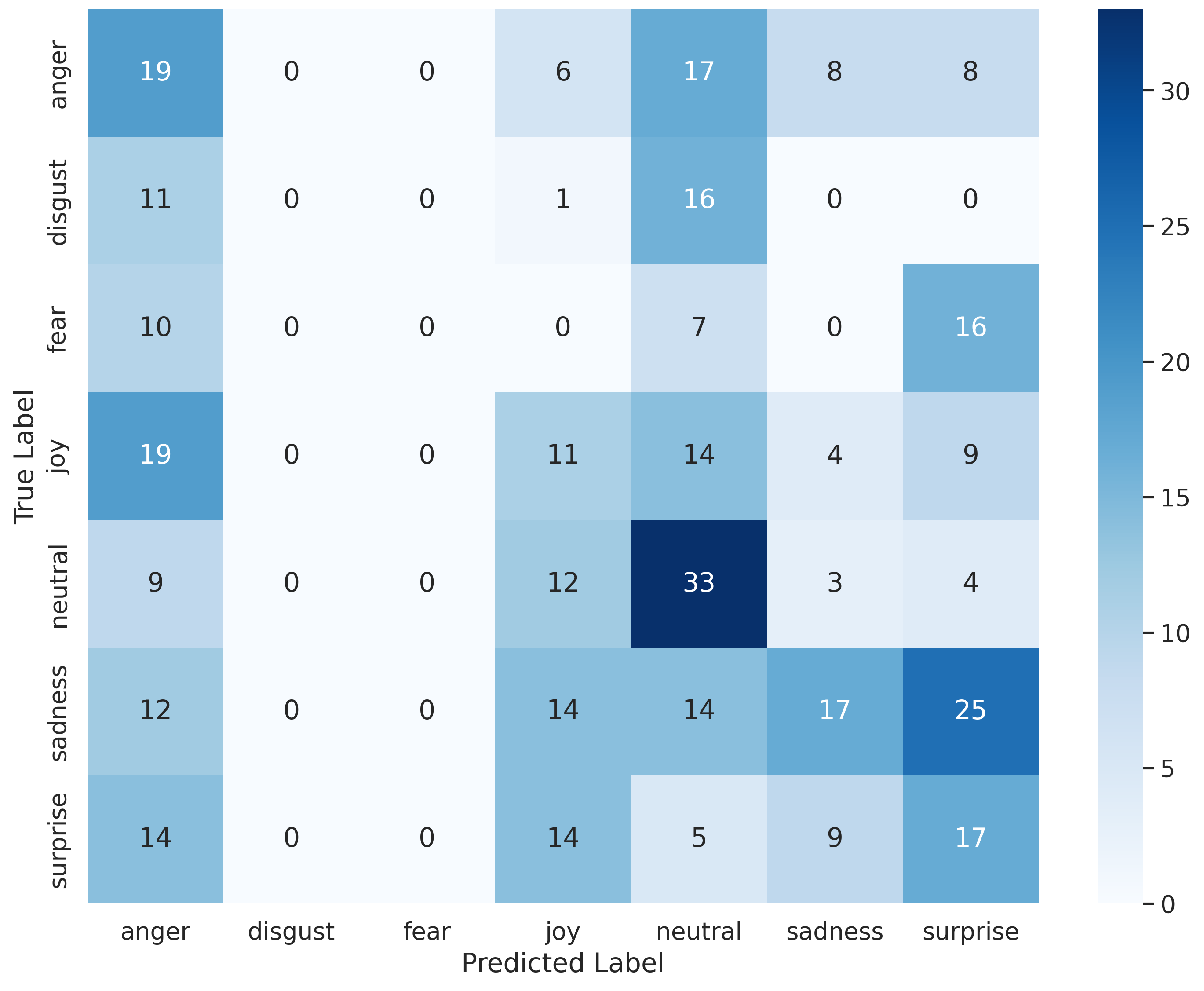}
        \caption{Wav2Vec 2.0 }
        \label{fig:CM_CASE_2c}
    \end{subfigure}

    \caption{Confusion matrices on our TWIN-SER dataset. Our DAS framework achieves the highest accuracy on major classes (\textit{anger}, \textit{sadness}, \textit{neutral}) and shows minimal confusion between high-arousal emotions. 
    }
    \label{fig:Confusion_Matrix_CASE}
\end{figure*}

To qualitatively assess the effectiveness of our DAS framework in learning discriminative emotion representations, we visualize the utterance-level embeddings using Uniform Manifold Approximation and Projection (UMAP)~\cite{UMAP}. Features are extracted from the final hidden layer (before the classifier) of each model and projected onto a 2D space for comparison on the TWIN-SER benchmark.

As shown in Figure~\ref{fig:umap_CASE}, the embedding distributions reveal clear differences in the representational capacity of each model under tone-word conflict scenarios. For DAS (ours), the projected embeddings form reasonably separated clusters corresponding to distinct emotion categories, indicating that our disentangled acoustic-semantic fusion effectively captures discriminative emotional cues despite the inherent conflicts between prosody and lexical content.  In contrast, the other baselines exhibit varying degrees of cluster overlap and dispersion. Whisper, though achieving the best performance among existing methods on TWIN-SER (47.26\% WA), still shows noticeable inter-category confusion, with several emotion classes (e.g., fear, sadness, surprise) partially overlapping. This reflects Whisper's reliance on semantic features while struggling to reconcile contradictory prosodic information. WavLM and Wav2Vec 2.0 yield even more scattered and overlapping clusters, consistent with their lower recognition accuracy. MingTok-Audio and VibeVoice, as acoustic-only tokenizers, produce embeddings with minimal category separation, confirming that acoustic features alone are insufficient for distinguishing emotions when lexical cues are misleading. The quantitative confusion patterns from the figure align with these qualitative observations, further corroborating that DAS achieves superior class separability on the challenging TWIN-SER benchmark. %Overall, the UMAP visualization provides intuitive evidence that our proposed DAS framework learns more robust and discriminative emotion representations, effectively mitigating the interference caused by tone-word conflicts.

\textbf{Confusion Matrix Analysis.}
To further evaluate the performance of our proposed DAS framework, we compare its confusion matrix against those of five representative baselines: MingTok-Audio, VibeVoice, HuBERT, WavLM, and Wav2Vec 2.0. As shown in Figure~\ref{fig:Confusion_Matrix_CASE}, DAS achieves the highest accuracy on major classes—particularly \textit{anger}, \textit{sadness}, and \textit{neutral}—with significantly less confusion between high-arousal emotions (e.g., \textit{anger} vs. \textit{surprise}) than other models. For instance, DAS correctly classifies a substantial proportion of \textit{anger} and \textit{sadness} samples, whereas other models frequently misclassify them as \textit{surprise}, reflecting their difficulty in distinguishing high-arousal negative emotions under tone-word conflicts.

Among the baselines, HuBERT, WavLM, and Wav2Vec 2.0 exhibit similarly scattered prediction patterns, frequently misclassifying samples as \textit{anger} or \textit{surprise}. This suggests that their self-supervised representations are largely shaped by acoustic features, which can override lexical-semantic cues and cause misalignment with the ground-truth emotions when tone-word conflicts occur. In contrast, the acoustic-only models—MingTok-Audio and VibeVoice—deliver consistently poor performance across all categories, further highlighting the limitations of relying exclusively on prosodic information for robust emotion recognition.

Notably, no method correctly predicts any samples of \textit{fear} or \textit{disgust}. This is likely due to two contributing factors: first, the TWIN-SER dataset contains a limited number of samples for these categories (only 33 and 28 samples, respectively), making it inherently difficult for any model to learn robust representations; second, both \textit{fear} and \textit{disgust} are intrinsically subtle emotions whose acoustic expression overlaps with \textit{sadness} and \textit{anger}, and this overlap is exacerbated in tone-word conflict scenarios where the lexical content may suggest a completely different emotion. The inability of all models to correctly classify these categories highlights the extreme difficulty of the TWIN-SER benchmark and underscores the need for more sophisticated approaches to handle such challenging minority classes. Overall, the confusion matrix comparison confirms that DAS effectively mitigates the interference caused by tone-word conflicts, achieving superior class-wise discrimination while maintaining balanced performance across categories.

\section{Conclusion}
\label{sec:conclusion}
In this paper, we tackle the acoustic-semantic conflict in SER, where prosody and lexical meaning convey contradictory emotions. We show that existing SER models are vulnerable to such conflicts due to semantic bias or entangled representations. To address this, we propose the DAS framework, which disentangles and bridges acoustic and semantic pathways via query-based fusion, along with the TWIN-SER benchmark for conflict robustness evaluation. Extensive experiments confirm that FAS outperforms state-of-the-art methods under both in-domain and zero-shot conditions. Future work may explore its integration into end-to-end audio-language models.

\textbf{Limitations and further work.} 
Despite the promising results, several limitations of this work should be acknowledged, which also point to promising directions for future research. While TWIN-SER provides a controlled environment for evaluating tone-word conflicts, it currently covers only two languages (English and Chinese). The dataset size (378 samples) is relatively modest compared to large-scale SER corpora, which may limit the statistical power for fine-grained analysis, particularly for rare emotion classes such as \textit{fear} and \textit{disgust}. Future work will expand TWIN-SER to include additional languages, more diverse speaking styles, and a broader range of emotion categories, as well as increase the total number of samples to support more robust model training and evaluation. DAS is currently implemented as a lightweight standalone module operating on pre-extracted features. Its potential as an integrated component within end-to-end audio-language models (ALMs) remains unexplored. Future work could investigate end-to-end fine-tuning of DAS with large multimodal backbones, which may further enhance its representational capacity and allow for more seamless interaction between acoustic and semantic pathways.

% \section*{Acknowledgments}
% This should be a simple paragraph before the References to thank those individuals and institutions who have supported your work on this article.


% \section{References}

% You can manually copy in the resultant .bbl file and set second argument of $\backslash${\tt{begin}} to the number of references
%  (used to reserve space for the reference number labels box).

% ---- References ----
\bibliographystyle{IEEEtran}
\bibliography{ref}

@article{abdollahi2022artificial,
  title={Artificial emotional intelligence in socially assistive robots for older adults: a pilot study},
  author={Abdollahi, Hojjat and Mahoor, Mohammad H and Zandie, Rohola and Siewierski, Jarid and Qualls, Sara H},
  journal={IEEE transactions on affective computing},
  volume={14},
  number={3},
  pages={2020--2032},
  year={2022},
  publisher={IEEE}
}

@ARTICLE{9543566,
  author={Latif, Siddique and Rana, Rajib and Khalifa, Sara and Jurdak, Raja and Qadir, Junaid and Schuller, Björn},
  journal={IEEE Transactions on Affective Computing}, 
  title={Survey of Deep Representation Learning for Speech Emotion Recognition}, 
  year={2023},
  volume={14},
  number={2},
  pages={1634-1654},
  doi={10.1109/TAFFC.2021.3114365}}

@article{wani2021comprehensive,
  title={A comprehensive review of speech emotion recognition systems},
  author={Wani, Taiba Majid and Gunawan, Teddy Surya and Qadri, Syed Asif Ahmad and Kartiwi, Mira and Ambikairajah, Eliathamby},
  journal={IEEE access},
  volume={9},
  pages={47795--47814},
  year={2021},
  publisher={IEEE}
}

@INPROCEEDINGS{10887715,
  author={Kwok, Chin Yuen and Li, Sheng and Yip, Jia Qi and Chu, Chenhui and Kawahara, Tatsuya and Chng, Eng Siong},
  booktitle={ICASSP 2025 - 2025 IEEE International Conference on Acoustics, Speech and Signal Processing (ICASSP)}, 
  title={Extending Whisper for Emotion Prediction Using Word-level Pseudo Labels}, 
  year={2025},
  volume={},
  number={},
  pages={1-5},
  doi={10.1109/ICASSP49660.2025.10887715}}

@INPROCEEDINGS{10446997,
  author={Goron, Erik and Asai, Lena and Rut, Elias and Dinov, Martin},
  booktitle={ICASSP 2024 - 2024 IEEE International Conference on Acoustics, Speech and Signal Processing (ICASSP)}, 
  title={Improving Domain Generalization in Speech Emotion Recognition with Whisper}, 
  year={2024},
  volume={},
  number={},
  pages={11631-11635},
  doi={10.1109/ICASSP48485.2024.10446997}}

@article{DBLP:journals/corr/abs-2111-02735,
  author       = {Yingzhi Wang and
                  Abdelmoumene Boumadane and
                  Abdelwahab Heba},
  title        = {A Fine-tuned Wav2vec 2.0/HuBERT Benchmark For Speech Emotion Recognition,
                  Speaker Verification and Spoken Language Understanding},
  journal      = {CoRR},
  volume       = {abs/2111.02735},
  year         = {2021},
  url          = {https://arxiv.org/abs/2111.02735},
  eprinttype   = {arXiv},
  eprint       = {2111.02735},
  bibsource    = {dblp computer science bibliography, https://dblp.org}
}

@article{jafarzadeh2024speaker,
  title={Speaker emotion recognition: Leveraging self-supervised models for feature extraction using Wav2Vec2 and HuBERT},
  author={Jafarzadeh, Pourya and Rostami, Amir Mohammad and Choobdar, Padideh},
  journal={arXiv preprint arXiv:2411.02964},
  year={2024}
}

@inproceedings{whisper,
  title={Robust speech recognition via large-scale weak supervision},
  author={Radford, Alec and Kim, Jong Wook and Xu, Tao and Brockman, Greg and McLeavey, Christine and Sutskever, Ilya},
  booktitle={International conference on machine learning},
  pages={28492--28518},
  year={2023},
  organization={PMLR}
}

@article{hubert,
  title={Hubert: Self-supervised speech representation learning by masked prediction of hidden units},
  author={Hsu, Wei-Ning and Bolte, Benjamin and Tsai, Yao-Hung Hubert and Lakhotia, Kushal and Salakhutdinov, Ruslan and Mohamed, Abdelrahman},
  journal={IEEE/ACM transactions on audio, speech, and language processing},
  volume={29},
  pages={3451--3460},
  year={2021},
  publisher={IEEE}
}

@article{IEMOCAP,
  title={IEMOCAP: Interactive emotional dyadic motion capture database},
  author={Busso, Carlos and Bulut, Murtaza and Lee, Chi-Chun and Kazemzadeh, Abe and Mower, Emily and Kim, Samuel and Chang, Jeannette N and Lee, Sungbok and Narayanan, Shrikanth S},
  journal={Language resources and evaluation},
  volume={42},
  number={4},
  pages={335--359},
  year={2008},
  publisher={Springer}
}

@inproceedings{MELD,
  title={Meld: A multimodal multi-party dataset for emotion recognition in conversations},
  author={Poria, Soujanya and Hazarika, Devamanyu and Majumder, Navonil and Naik, Gautam and Cambria, Erik and Mihalcea, Rada},
  booktitle={Proceedings of the 57th annual meeting of the association for computational linguistics},
  pages={527--536},
  year={2019}
}

@inproceedings{emotion2vec,
  title={emotion2vec: Self-supervised pre-training for speech emotion representation},
  author={Ma, Ziyang and Zheng, Zhisheng and Ye, Jiaxin and Li, Jinchao and Gao, Zhifu and Zhang, Shiliang and Chen, Xie},
  booktitle={Findings of the Association for Computational Linguistics: ACL 2024},
  pages={15747--15760},
  year={2024}
}

@article{C2SER,
  title={Steering Language Model to Stable Speech Emotion Recognition via Contextual Perception and Chain of Thought},
  author={Zhao, Zhixian and Zhu, Xinfa and Wang, Xinsheng and Wang, Shuiyuan and Geng, Xuelong and Tian, Wenjie and Xie, Lei},
  journal={arXiv preprint arXiv:2502.18186},
  year={2025}
}

@article{Xcodec2,
  title={Llasa: Scaling train-time and inference-time compute for llama-based speech synthesis},
  author={Ye, Zhen and Zhu, Xinfa and Chan, Chi-Min and Wang, Xinsheng and Tan, Xu and Lei, Jiahe and Peng, Yi and Liu, Haohe and Jin, Yizhu and Dai, Zheqi and others},
  journal={arXiv preprint arXiv:2502.04128},
  year={2025}
}

@inproceedings{Xcodec,
  title={Codec does matter: Exploring the semantic shortcoming of codec for audio language model},
  author={Ye, Zhen and Sun, Peiwen and Lei, Jiahe and Lin, Hongzhan and Tan, Xu and Dai, Zheqi and Kong, Qiuqiang and Chen, Jianyi and Pan, Jiahao and Liu, Qifeng and others},
  booktitle={Proceedings of the AAAI Conference on Artificial Intelligence},
  volume={39},
  number={24},
  pages={25697--25705},
  year={2025}
}

@misc{VibeVoice,
      title={VibeVoice Technical Report}, 
      author={Zhiliang Peng and Jianwei Yu and Wenhui Wang and Yaoyao Chang and Yutao Sun and Li Dong and Yi Zhu and Weijiang Xu and Hangbo Bao and Zehua Wang and Shaohan Huang and Yan Xia and Furu Wei},
      year={2025},
      eprint={2508.19205},
      archivePrefix={arXiv},
      primaryClass={cs.CL},
      url={https://arxiv.org/abs/2508.19205}, 
}

@article{Ming-UniAudio,
  title={Ming-UniAudio: Speech LLM for Joint Understanding, Generation and Editing with Unified Representation},
  author={Yan, Canxiang and Jin, Chunxiang and Huang, Dawei and Yu, Haibing and Peng, Han and Zhan, Hui and Gao, Jie and Peng, Jing and Chen, Jingdong and Zhou, Jun and others},
  journal={arXiv preprint arXiv:2511.05516},
  year={2025}
}

@misc{Encodec,
      title={High Fidelity Neural Audio Compression}, 
      author={Alexandre Défossez and Jade Copet and Gabriel Synnaeve and Yossi Adi},
      year={2022},
      eprint={2210.13438},
      archivePrefix={arXiv},
      primaryClass={eess.AS},
      url={https://arxiv.org/abs/2210.13438}, 
}

@INPROCEEDINGS{CLAP,
  author={Elizalde, Benjamin and Deshmukh, Soham and Ismail, Mahmoud Al and Wang, Huaming},
  booktitle={ICASSP 2023 - 2023 IEEE International Conference on Acoustics, Speech and Signal Processing (ICASSP)}, 
  title={CLAP Learning Audio Concepts from Natural Language Supervision}, 
  year={2023},
  pages={1-5},
  doi={10.1109/ICASSP49357.2023.10095889}
}

@article{RAVDESS,
  title={The Ryerson Audio-Visual Database of Emotional Speech and Song (RAVDESS): A dynamic, multimodal set of facial and vocal expressions in North American English},
  author={Livingstone, Steven R and Russo, Frank A},
  journal={PloS one},
  volume={13},
  number={5},
  pages={e0196391},
  year={2018},
  publisher={Public Library of Science San Francisco, CA USA}
}

@article{ESD,
title={Emotional voice conversion: Theory, databases and ESD},
author={Zhou, Kun and Sisman, Berrak and Liu, Rui and Li, Haizhou},
journal={Speech Communication},
volume={137},
pages={1--18},
year={2022},
publisher={Elsevier}
}

@inproceedings{EmoDB,
  title={A database of german emotional speech.},
  author={Burkhardt, Felix and Paeschke, Astrid and Rolfes, Miriam and Sendlmeier, Walter F and Weiss, Benjamin and others},
  booktitle={Interspeech},
  volume={5},
  pages={1517--1520},
  year={2005}
}

@article{wav2vec,
  title={wav2vec: Unsupervised pre-training for speech recognition},
  author={Schneider, Steffen and Baevski, Alexei and Collobert, Ronan and Auli, Michael},
  journal={arXiv preprint arXiv:1904.05862},
  year={2019}
}

@article{wav2vec2.0,
  title={wav2vec 2.0: A framework for self-supervised learning of speech representations},
  author={Baevski, Alexei and Zhou, Yuhao and Mohamed, Abdelrahman and Auli, Michael},
  journal={Advances in neural information processing systems},
  volume={33},
  pages={12449--12460},
  year={2020}
}

@article{Wavlm,
  title={Wavlm: Large-scale self-supervised pre-training for full stack speech processing},
  author={Chen, Sanyuan and Wang, Chengyi and Chen, Zhengyang and Wu, Yu and Liu, Shujie and Chen, Zhuo and Li, Jinyu and Kanda, Naoyuki and Yoshioka, Takuya and Xiao, Xiong and others},
  journal={IEEE Journal of Selected Topics in Signal Processing},
  volume={16},
  number={6},
  pages={1505--1518},
  year={2022},
  publisher={IEEE}
}

@article{Vesper,
  title={Vesper: A compact and effective pretrained model for speech emotion recognition},
  author={Chen, Weidong and Xing, Xiaofen and Chen, Peihao and Xu, Xiangmin},
  journal={IEEE Transactions on Affective Computing},
  volume={15},
  number={3},
  pages={1711--1724},
  year={2024},
  publisher={IEEE}
}

@article{MFGCN,
title = {MFGCN: Multimodal fusion graph convolutional network for speech emotion recognition},
journal = {Neurocomputing},
volume = {611},
pages = {128646},
year = {2025},
issn = {0925-2312},
doi = {https://doi.org/10.1016/j.neucom.2024.128646},
url = {https://www.sciencedirect.com/science/article/pii/S0925231224014176},
author = {Xin Qi and Yujun Wen and Pengzhou Zhang and Heyan Huang}
}

@article{VQ-VAE,
  title={Neural discrete representation learning},
  author={Van Den Oord, Aaron and Vinyals, Oriol and others},
  journal={Advances in neural information processing systems},
  volume={30},
  year={2017}
}

@InProceedings{DiTAR,
  title = 	 {{D}i{TAR}: Diffusion Transformer Autoregressive Modeling for Speech Generation},
  author =       {Jia, Dongya and Chen, Zhuo and Chen, Jiawei and Du, Chenpeng and Wu, Jian and Cong, Jian and Zhuang, Xiaobin and Li, Chumin and Wei, Zhen and Wang, Yuping and Wang, Yuxuan},
  booktitle = 	 {Proceedings of the 42nd International Conference on Machine Learning},
  pages = 	 {27255--27270},
  year = 	 {2025},
  volume = 	 {267},
  series = 	 {Proceedings of Machine Learning Research},
  month = 	 {13--19 Jul},
  publisher =    {PMLR},
  url = 	 {https://proceedings.mlr.press/v267/jia25a.html},
}

@inproceedings{CMU-MOSEI,
    title = "Multimodal Language Analysis in the Wild: {CMU}-{MOSEI} Dataset and Interpretable Dynamic Fusion Graph",
    author = "Bagher Zadeh, AmirAli  and
      Liang, Paul Pu  and
      Poria, Soujanya  and
      Cambria, Erik  and
      Morency, Louis-Philippe",
    booktitle = "Proceedings of the 56th Annual Meeting of the Association for Computational Linguistics (Volume 1: Long Papers)",
    month = jul,
    year = "2018",
    address = "Melbourne, Australia",
    publisher = "Association for Computational Linguistics",
    url = "https://aclanthology.org/P18-1208/",
    doi = "10.18653/v1/P18-1208",
    pages = "2236--2246"
}

@misc{MER2024,
      title={MER 2024: Semi-Supervised Learning, Noise Robustness, and Open-Vocabulary Multimodal Emotion Recognition}, 
      author={Zheng Lian and Haiyang Sun and Licai Sun and Zhuofan Wen and Siyuan Zhang and Shun Chen and Hao Gu and Jinming Zhao and Ziyang Ma and Xie Chen and Jiangyan Yi and Rui Liu and Kele Xu and Bin Liu and Erik Cambria and Guoying Zhao and Björn W. Schuller and Jianhua Tao},
      year={2024},
      eprint={2404.17113},
      archivePrefix={arXiv},
      primaryClass={cs.LG},
      url={https://arxiv.org/abs/2404.17113}, 
}

@inproceedings{BLIP-2,
  title={Blip-2: Bootstrapping language-image pre-training with frozen image encoders and large language models},
  author={Li, Junnan and Li, Dongxu and Savarese, Silvio and Hoi, Steven},
  booktitle={International conference on machine learning},
  pages={19730--19742},
  year={2023},
  organization={PMLR}
}

@inproceedings{CLIP,
  title={Learning transferable visual models from natural language supervision},
  author={Radford, Alec and Kim, Jong Wook and Hallacy, Chris and Ramesh, Aditya and Goh, Gabriel and Agarwal, Sandhini and Sastry, Girish and Askell, Amanda and Mishkin, Pamela and Clark, Jack and others},
  booktitle={International conference on machine learning},
  pages={8748--8763},
  year={2021},
  organization={PmLR}
}

@article{li2025seenet,
  title={SeeNet: A soft emotion expert and data augmentation method to enhance speech emotion recognition},
  author={Li, Qifei and Gao, Yingming and Wen, Yuhua and Zhao, Ziping and Li, Ya and Schuller, Bj{\"o}rn W},
  journal={IEEE Transactions on Affective Computing},
  year={2025},
  publisher={IEEE}
}

@INPROCEEDINGS{10890265,
  author={Zhao, Jiaqi and Wang, Fei and Li, Kun and Wei, Yanyan and Tang, Shengeng and Zhao, Shu and Sun, Xiao},
  booktitle={ICASSP 2025 - 2025 IEEE International Conference on Acoustics, Speech and Signal Processing (ICASSP)}, 
  title={Temporal-Frequency State Space Duality: An Efficient Paradigm for Speech Emotion Recognition}, 
  year={2025},
  volume={},
  number={},
  pages={1-5},
  doi={10.1109/ICASSP49660.2025.10890265}}

@article{Seed-tts,
  title={Seed-tts: A family of high-quality versatile speech generation models},
  author={Anastassiou, Philip and Chen, Jiawei and Chen, Jitong and Chen, Yuanzhe and Chen, Zhuo and Chen, Ziyi and Cong, Jian and Deng, Lelai and Ding, Chuang and Gao, Lu and others},
  journal={arXiv preprint arXiv:2406.02430},
  year={2024}
}

@article{Gemini2.5,
  title={Gemini 2.5: Pushing the frontier with advanced reasoning, multimodality, long context, and next generation agentic capabilities},
  author={Comanici, Gheorghe and Bieber, Eric and Schaekermann, Mike and Pasupat, Ice and Sachdeva, Noveen and Dhillon, Inderjit and Blistein, Marcel and Ram, Ori and Zhang, Dan and Rosen, Evan and others},
  journal={arXiv preprint arXiv:2507.06261},
  year={2025}
}

@article{AdamW,
  title={Decoupled weight decay regularization},
  author={Loshchilov, Ilya and Hutter, Frank},
  journal={arXiv preprint arXiv:1711.05101},
  year={2017}
}

@article{UMAP,
  title={Umap: Uniform manifold approximation and projection for dimension reduction},
  author={McInnes, Leland and Healy, John and Melville, James},
  journal={arXiv preprint arXiv:1802.03426},
  year={2018}
}

@article{qwen2_audio,
  title={Qwen2-audio technical report},
  author={Chu, Yunfei and Xu, Jin and Yang, Qian and Wei, Haojie and Wei, Xipin and Guo, Zhifang and Leng, Yichong and Lv, Yuanjun and He, Jinzheng and Lin, Junyang and others},
  journal={arXiv preprint arXiv:2407.10759},
  year={2024}
}

@article{Qwen2_5_omni,
  title={Qwen2. 5-omni technical report},
  author={Xu, Jin and Guo, Zhifang and He, Jinzheng and Hu, Hangrui and He, Ting and Bai, Shuai and Chen, Keqin and Wang, Jialin and Fan, Yang and Dang, Kai and others},
  journal={arXiv preprint arXiv:2503.20215},
  year={2025}
}

@inproceedings{opensmile,
  title={Recent developments in opensmile, the munich open-source multimedia feature extractor},
  author={Eyben, Florian and Weninger, Felix and Gross, Florian and Schuller, Bj{\"o}rn},
  booktitle={Proceedings of the 21st ACM international conference on Multimedia},
  pages={835--838},
  year={2013}
}

@article{Survey_on_SER,
  title={Survey on speech emotion recognition: Features, classification schemes, and databases},
  author={El Ayadi, Moataz and Kamel, Mohamed S and Karray, Fakhri},
  journal={Pattern recognition},
  volume={44},
  number={3},
  pages={572--587},
  year={2011},
  publisher={Elsevier}
}

@article{SER_CNN,
  title={Speech emotion recognition using convolutional neural networks with attention mechanism},
  author={Mountzouris, Konstantinos and Perikos, Isidoros and Hatzilygeroudis, Ioannis},
  journal={Electronics},
  volume={12},
  number={20},
  pages={4376},
  year={2023},
  publisher={MDPI}
}

@article{SER_1D2D,
  title={Speech emotion recognition using deep 1D \& 2D CNN LSTM networks},
  author={Zhao, Jianfeng and Mao, Xia and Chen, Lijiang},
  journal={Biomedical signal processing and control},
  volume={47},
  pages={312--323},
  year={2019},
  publisher={Elsevier}
}

@inproceedings{MUSTARD,
  title={Towards multimodal sarcasm detection (an \_obviously\_ perfect paper)},
  author={Castro, Santiago and Hazarika, Devamanyu and P{\'e}rez-Rosas, Ver{\'o}nica and Zimmermann, Roger and Mihalcea, Rada and Poria, Soujanya},
  booktitle={Proceedings of the 57th annual meeting of the association for computational linguistics},
  pages={4619--4629},
  year={2019}
}

@inproceedings{iSarcasm/iSarcasmEval,
  title={Semeval-2022 task 6: isarcasmeval, intended sarcasm detection in english and arabic},
  author={Farha, Ibrahim Abu and Oprea, Silviu Vlad and Wilson, Steven and Magdy, Walid},
  booktitle={Proceedings of the 16th International Workshop on Semantic Evaluation (SemEval-2022)},
  pages={802--814},
  year={2022}
}

% \newpage
% \vspace{-10mm}

 

% \end{IEEEbiography}

\vspace{11pt}


% \newpage
\vspace{-12mm}
\begin{IEEEbiography}
[{\includegraphics[width=1in,height=1.25in,clip,keepaspectratio]{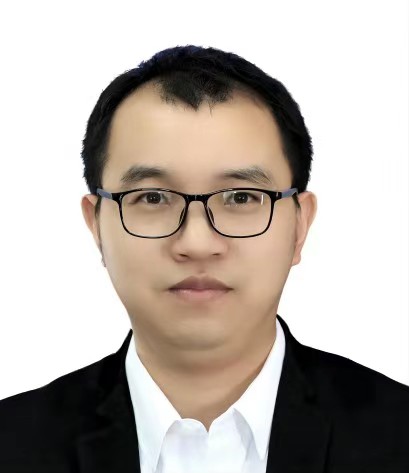}}]{Xiaojiang Peng}
(Senior Member, IEEE) received his Ph.D. degree in Computer Science from Southwest Jiaotong University, China, in 2015. Currently, he is a full professor at Shenzhen Technology University and serves as the dean of the Artificial Intelligence Department. He was an associate professor
at the Chinese Academy of Sciences and a postdoctoral researcher
with Idiap, Switzerland, and Inria THOTH, France. He has published more than 100 top journal/conference papers, and serves as an associate editor for IEEE Transactions on Affective Computing, and a reviewer for the International Journal of Computer Vision, IEEE Transactions on
Pattern Analysis and Machine Intelligence, IEEE Transactions on Image Processing, CVPR, ICCV, ECCV, AAAI, IJCAI, FG, etc. He has been selected by Stanford University as one of the world's top 2\% scientists
in 2022, 2023, 2024, and 2025. His research interests include computer vision, affective computing, embodied intelligence, and generative models.
\end{IEEEbiography}
% \vspace{-20\baselineskip} % 减少上半行距
 \vspace{-10mm}

\begin{IEEEbiography}[{\includegraphics[width=1in,height=1.25in,clip,keepaspectratio]{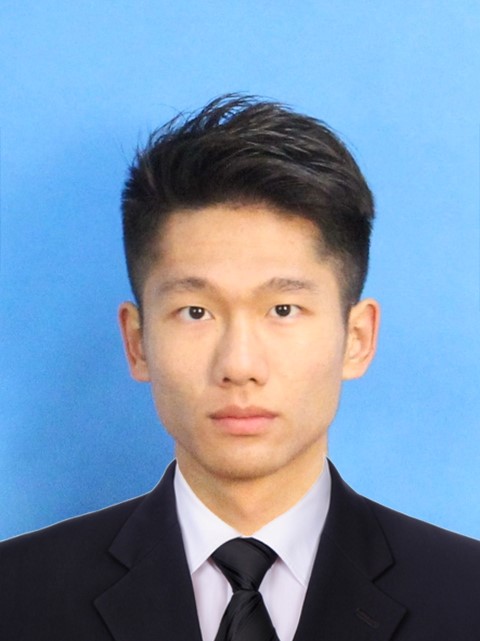}}]{Dawei Huang} received the B.S. degree in Mathematics from the University of Science and Technology Beijing, China. He is currently pursuing the M.S. degree in Computer Science and Technology at Shenzhen University, China. He is also a Research Intern with the Bailing Team for Large Multimodal Models at Ant Group. His research interests include multimodal emotion recognition, large multimodal models, and text-to-speech synthesis.
\end{IEEEbiography}
 \vspace{-10mm}

\begin{IEEEbiographynophoto}{Yongjie Lv} received the M.S. degree from Tianjin University, Tianjin, China, in 2022. He is currently a Researcher at Ant Group, Hangzhou, China. His research interests include speech understanding and generation.
\end{IEEEbiographynophoto}
 \vspace{-10mm}
 
\begin{IEEEbiographynophoto}{Ruijie Xiong} received the M.S. degree from Harbin Institute of Technology, Harbin, China, in 2021. He was a Principal Researcher with Tencent AI Lab from 2021 to 2024. He is currently a Principal Researcher at Alibaba Group, Hangzhou, China. His research interests focus on speech processing and related multimodal technologies.
\end{IEEEbiographynophoto}
 \vspace{-10mm}
 
\begin{IEEEbiographynophoto}{Chunxiang Jin} received the M.S. degree from Shanghai Institute of Optics and Fine Mechanics, University of Chinese Academy of Sciences, Shanghai, China, in 2015. He is currently a Principal Researcher and leads the Speech Multimodal Team within the Foundation Model Department at Ant Group, Hangzhou, China, where he has been since 2017. His research focuses on speech foundation models, including Ming-UniAudio and Ming-Omni-TTS.
\end{IEEEbiographynophoto}
 \vspace{-10mm}

\begin{IEEEbiographynophoto}{Bin Li} received the degree in mechanical engineering from Xi’an Jiaotong University in 2002. He is currently the Deputy General Manager of the R\&D Center at Shenzhen Skyworth Digital Technology Co., Ltd.
\end{IEEEbiographynophoto}
 \vspace{-10mm}

\begin{IEEEbiographynophoto}{Xiaohui Wang} received the Ph.D. degree in communication engineering from Huazhong University of Science and Technology in 2000. He is currently the General Manager of Shenzhen Xiaopai Technology Co., Ltd. 
\end{IEEEbiographynophoto}
 \vspace{-10mm}
 
\begin{IEEEbiography}[{\includegraphics[width=1in,height=1.25in,clip,keepaspectratio]{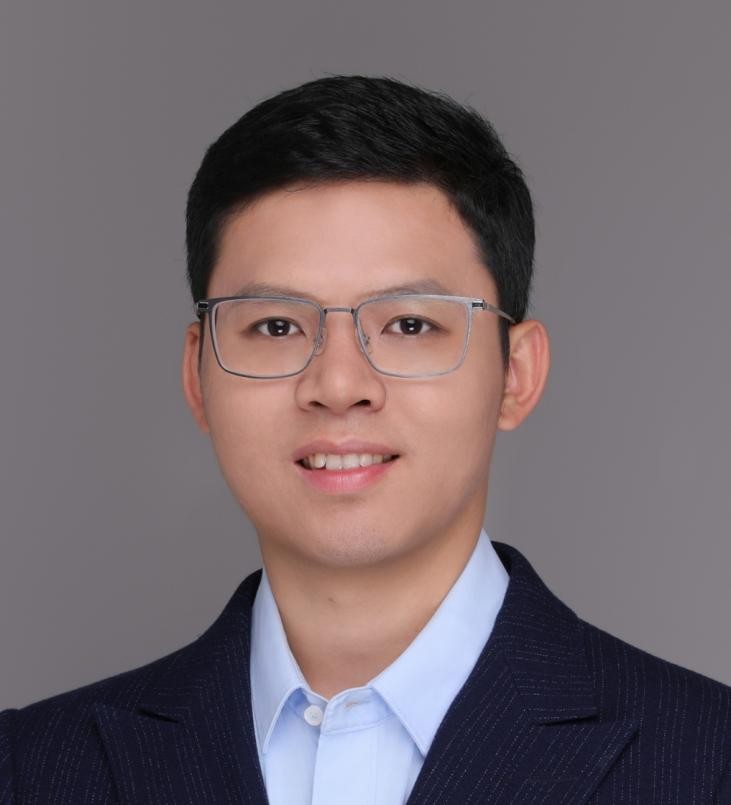}}]{Zitong Yu} (Senior Member, IEEE) received the Ph.D. degree in Computer Science and Engineering from the University of Oulu, Finland, in 2022. Currently, he is an Associate Professor (Tenured) at Great Bay University, China. He was a Postdoctoral researcher at ROSE Lab, Nanyang Technological University. He was a visiting scholar at TVG, University of Oxford, from July to November 2021. His research interests focus on subtle visual computing. He was a recipient of IAPR Best Student Paper Award, IEEE Finland Section Best Student Conference Paper Award. 
\end{IEEEbiography}
 \vspace{-10mm}
 
\vfill

\end{document}